\documentclass[acmsmall,authorversion,nonacm,screen]{acmart}
\usepackage{amsmath}
\usepackage[utf8]{inputenc}
\usepackage{color, soul}
\usepackage{physics}
\usepackage[ruled,vlined]{algorithm2e}
\usepackage{listings}
\usepackage{algorithmic}

\setcopyright{none}
\copyrightyear{}
\acmYear{2021}
\acmDOI{}

\acmConference[]{}{}{}
\acmBooktitle{}
\acmPrice{}
\acmISBN{}

\usepackage{subcaption}
\usepackage{graphicx}
\usepackage{savesym}
\savesymbol{eval}
\usepackage{semantic}
\restoresymbol{SEM}{eval}
\newcommand{\freezecache}{frozencache,cachedir=styminted}

\usepackage[\freezecache]{minted}

\usepackage{tikz}
\usepackage{pgf}
\usetikzlibrary{quantikz}

\pgfdeclarelayer{nodelayer}
\pgfdeclarelayer{edgelayer}
\pgfsetlayers{main,nodelayer,edgelayer}

\definecolor{zx_red}{RGB}{232, 165, 165}
\definecolor{zx_green}{RGB}{216, 248, 216}

\tikzset{gn/.style={rectangle,rounded corners=0.8em,fill=zx_green,draw=black,
  line width=0.8 pt,inner sep=3pt,minimum width=1.5em,minimum height=1.5em}}
\tikzset{rn/.style={rectangle,rounded corners=0.8em,fill=zx_red,draw=black,
  line width=0.8 pt,inner sep=3pt,minimum width=1.5em,minimum height=1.5em}}
\tikzset{had/.style={fill=yellow, draw=black, shape=rectangle,line width=0.8 pt,inner sep=3pt,minimum width=.5em,minimum height=.5em}}

\newcommand{\assert}[1]{\{#1\}}
\newcommand{\hoare}[3]{\assert{#1} #2 \assert{#3}}

\newcommand{\change}{\textcolor{black}}
\def\eg{{\em e.g., }}

\def\Markov{{M}arkov }
\def\Hamiltonian{{H}amiltonian }
\def\Hamiltonians{{H}amiltonians }

\begin{document}
\title[Formal Verification of Quantum Programs]{Formal Verification of Quantum Programs: Theory, Tools and Challenges}

\author{Marco Lewis}
\email{m.j.lewis2@newcastle.ac.uk}
\author{Sadegh Soudjani}
\email{sadegh.soudjani@newcastle.ac.uk}
\author{Paolo Zuliani}
\email{paolo.zuliani@newcastle.ac.uk}
\affiliation{
    \institution{Newcastle University}
    \city{Newcastle-upon-Tyne}
    \country{UK}
}

\renewcommand{\shortauthors}{Lewis, Soudjani, Zuliani}

\begin{abstract}
Over the past 27 years, quantum computing has seen a huge rise in interest from both academia and industry. At the current rate, quantum computers are growing in size rapidly backed up by the increase of research in the field. Significant efforts are being made to improve the reliability of quantum hardware and to develop suitable software to program quantum computers.

In contrast, the verification of quantum programs has received relatively less attention. Verifying programs is especially important in the quantum setting due to how difficult it is to program complex algorithms correctly on resource-constrained and error-prone quantum hardware. Research into creating verification frameworks for quantum programs has seen recent development, with a variety of tools implemented using a collection of theoretical ideas.

This survey aims to be a short introduction into the area of formal verification of quantum programs, bringing together theory and tools developed to date. Further, this survey examines some of the challenges that the field may face in the future, namely the development of complex quantum algorithms.
\end{abstract}

\maketitle

\section{Introduction}
Quantum computers are capable of solving a variety of problems much faster than classical computers, as seen in Deutsch-Jozsa's and Grover's algorithms~\cite{Deutsch1992, Grover96}. With recent results from Google~\cite{Arute19} and the Hefei National Laboratory
\cite{Zhong20, zhu2021}, it is becoming more realistic that large-scale quantum computers will be used by academics and companies alike. Within the next few years, we expect to see a quantum computer with the ability to perform computations on 100s or potentially 1{,}000 quantum bits (otherwise known as qubits), with IBM aiming to debut a 1{,}121 qubit computer in 2023.\footnote{
\url{https://www.research.ibm.com/blog/quantum-development-roadmap/}
}
These planned devices are by no means ideal, as they are Noisy Intermediate-Scale Quantum (NISQ)~\cite{NISQ} computers that are limited by having hundreds of qubits but no error correction. The quantum threshold theorem~\cite{QTT} provides the theoretical justification as for why arbitrarily long quantum computations are possible in principle, albeit it assumes quantum error correction.

The problem of error in quantum computers needs to be solved to make large-scale quantum computers usable. There are three potential reasons why an incorrect value is measured during a computation on a quantum circuit. 
The first is due to the innate randomness when measuring qubits within quantum computers: {When a qubit is measured, it commonly collapses into one of its two `basis' states.}
Most quantum algorithms will only return the correct result with high probability {(for example, consider the probability that Grover's algorithm will return the marked element of a database)}. Therefore, there is still a chance of measuring incorrect results, but this can be solved by running a quantum circuit multiple times and returning the results that is measured the most (as the correct result will be measured most of the time).

Another source of error is the hardware. Hardware error occurs when qubits are interfered with by gates or sources from outside of the system. Qubits may change state when going through a gate in a different way to what is expected, often referred to as {gate infidelity}. 
Alternatively, errors can occur at readout when qubits are measured incorrectly. To prevent errors that change the phase or bit value of a qubit, error-correcting codes can be used~\cite{Roffe19} (see for example~\cite{Paler15} for a thorough investigation of the various techniques for achieving fault-tolerant quantum computing).

The final source of error is from within the software. This can occur at various stages, whether that would be when a programmer codes the algorithm incorrectly or if the compiled circuit is flawed.
To solve the error that comes within software, tools need to be developed to verify programs and compiled circuits.
By reducing errors within hardware and software, it can be guaranteed that readout distributions match those predicted by quantum theory.

{This article is focused on the error caused by software and how to prevent it through the usage of formal verification.} That involves the study and development of tools to verify programs and systems. Recent decades have seen the successful development of automatic provers capable of proving properties about programs or mathematical theories with the press of a button. {Examples of such tools}\footnote{Some can be found competing at competitions including SMT-COMP ({\url{https://smt-comp.github.io/}}) for SMT solvers and ARCH ({\url{https://cps-vo.org/group/ARCH}}) for hybrid \change{system} model checkers.}
include SMT (Satisfiability Modulo Theories) solvers: Z3~\cite{DeMoura08}, Yices2~\cite{Yices2} and dReal~\cite{dreal}; stochastic, discrete model checkers: PRISM~\cite{KNP11} and  STORM~\cite{STORM}; non-deterministic, discrete model checkers: nuSMV~\cite{nuSMV} and SPIN~\cite{Spin}; and stochastic, non-deterministic hybrid model checkers: ProbReach~\cite{probreach}, SpaceEx~\cite{SpaceEx} and Uppaal~\cite{Uppaal}.

Developing tools that ensure quantum software is correct is vital for quantum computers.
\change{
Quantum programs are hard to develop and even harder to test. Ensuring that a programmer is implementing an algorithm without any errors is particularly difficult in the quantum domain and requires dedicated tools to check for correctness. Further, pursuing research down this route may lead to better understanding of quantum computation and show the power of quantum computers over classical computers for certain tasks.
}

The goal of this {survey is to introduce the formal verification techniques and tools that have been developed to prevent errors and bugs within quantum programs.}
This includes summarising some of the theories that verification tools for quantum programs are based on and investigating implementations of these tools. Future hurdles for verifying complex quantum algorithms and routes for further research are also discussed.

Note that this article is mainly discussing techniques for verifying programs. Verification is important in other areas of computer science, such as communication, security, program equivalence and concurrency.
In this article, \change{there will be limited discussion on circuit equivalence (lower level verification of quantum circuits).}
\change{A recent survey} \cite{FMQCsurvey} provides an in-depth discussion on \change{circuit equivalence} and details on quantum programming languages. In contrast, our article focuses on designing verification frameworks and how more complex quantum algorithms need to be verified.

\change{Otherwise, readers are referred to~\cite{Unruh19, EasyQPC} for verification in quantum security, to~\cite{Davidson12} for verification in quantum communication, and to~\cite{Ardeshir18} for verification in concurrent quantum systems.
Process calculi/algebras lie at the intersection between communication and concurrency and have been extended to the quantum domain, allowing formal models to analyse quantum communication protocols. Such calculi include CQP~\cite{Gay2004}, QPAlg~\cite{Lalire2006} and qCCS~\cite{Ying2007}.}

The organisation of the paper is as follows.
Section~\ref{sec:bg} gives an overview of quantum computing and some background on the fundamentals of model checking and deductive verification.
Section~\ref{sec:qver} discusses the various formal verification techniques that have been extended or created to check quantum programs.
Section~\ref{sec:criteria} considers what is desirable in verifiable quantum programming languages.
Section~\ref{sec:vqpls} discusses the most recent verifiable tools for verifying quantum programs with relation to the topics discussed in the previous section.
Section~\ref{sec:alg} gives examples of non-textbook algorithms and details what hurdles need to be overcome to verify them.

\section{Background \label{sec:bg}}

This section introduces \change{the standard notation used in quantum computing, and} the field of formal verification.
Whilst there are many techniques for formal verification, this section focuses on the two most popular ones: model checking and deductive verification.

\subsection{Quantum Computing Notation \label{sec:bg:quantum}}
Nielsen and Chuang's volume \cite{Nielsen11} is the standard textbook for quantum computing \change{and a full introduction can be found therein. This section will briefly cover some notation used throughout the paper, however new notation is introduced where appropriate.}

\change{Throughout, we make use of the Dirac/bra-ket notation to describe quantum states and operations.}
\change{The states $\ket{0} = [1, 0]^\intercal $ and $\ket{1} = [0,1]^\intercal$ describe the computational basis states.
In general, a quantum state is described as $\ket{\phi} = \sum_j \alpha_j \ket{\phi_j}$ where $\phi_j$ is typically bitstring. The dual of a quantum state is denoted by a bra, $\ket{\phi}^\dag = \bra{\phi}$.}

\change{Unitary operations, denoted by $U$, are operations from quantum states to quantum states and their inverse is their adjoint. So, $U^{-1} = U^\dagger = \overline{U^\intercal}$. We use $U\ket{\phi}$ to mean the result of applying $U$ to $\ket{\phi}$.}

\change{Hermitian operators, denoted by $H$, are operators that are self-adjoint, so $H = H^\dagger$. Further, Hermitian operators have real eigenvalues.}
%
\change{The density matrix formalism is also discussed and used instead to represent quantum states. In this setting, states are described by Hermitian matrices and often are written as $\rho = \sum_j \alpha_j \ketbra{\phi_j}$. This representation is used in various parts of the paper, notably Sections~\ref{sec:qhl} and \ref{lang:QHLProver}.}

\subsection{Model Checking and Verification \label{sec:bg:model}}

Verifying software with model checking involves modelling the software through a formal representation. Then desired behaviour is specified through an appropriate logic. Once these two components are created, model checking algorithms can be used to check whether the model follows the specified behaviour. In this section, we study the temporal behaviour of software and systems. Models are created using Kripke structures and behaviours are specified using a temporal logic. More details can be found in~\cite{MCBook}.

Kripke structures model software or systems by describing transitions between states in a similar way to finite state machines. But Kripke structures also model properties that hold in each state. Formally:
\change{
\begin{definition} \label{def:kripke}
A Kripke structure is given by a 4-tuple $M = (S, S_0, R, L)$ where
\begin{itemize}
    \item $S$ is a finite set of states;
    \item $S_0 \subseteq S$ is \change{the set of} initial states;
    \item $R \subseteq S \times S$ is a total transition relation, where for all $s \in S$, there exists $s' \in S$ such that $(s, s') \in R$;
    \item $L: S \rightarrow 2^{AP}$ is a labelling function that gives the set of propositions ($p \in AP$) that hold within a given state.
\end{itemize}
\end{definition}
}

A common type of logic used to specify behaviour is temporal logic, which can be used to describe what propositions may hold about the system over time. Examples of temporal logics include Linear Temporal Logic (LTL) \mbox{\cite{LTL}}, Computation Tree Logic (CTL) \mbox{\cite{CTL}} and the $\mu$-calculus \cite{mu-calc}.
The definition of CTL is given and briefly studied as it will be useful for understanding Section~\ref{sec:qctl}.

\change{Before giving formal semantics of CTL, we first define the paths on a Kripke structure. Throughout, let ${M=(S, S_0, R, L)}$ be a Kripke structure.
\begin{definition}
A path is a tuple $\sigma = (s_0, s_1, s_2, \dots)$ where $s_i \in S$ and we have that $(s_i, s_{i+1}) \in R$.
\end{definition}
Note that a path can have infinite or finite length as long as there are suitable transitions.
}

Terms in CTL are given by state formulae, $\theta$, and temporal operators, $T$, that only exist bound with path quantifiers. These formulae are defined inductively by
\begin{align*}
    &\theta ::= p
    \text{ || }\lnot \theta
    \text{ || } \theta \lor \theta
    \text{ || } \textbf{E} T
    \text{ || } \textbf{A} T
    \\
    & T ::= \textbf{X} \theta
    \text{ || } \textbf{F} \theta
    \text{ || } \textbf{G} \theta
    \text{ || } \theta \textbf{U} \theta {,}
\end{align*}
where $p \in AP$ is an atomic proposition in the model. The terms \textbf{X} (``next''), \textbf{F} (``eventually''), \textbf{G} (``always'') and  \textbf{U} (``until'') denote basic temporal operations. The terms \textbf{A} (for all paths) and \textbf{E} (there exists a path) are path quantifiers.
\change{The semantics of the state and temporal operations described for a Kripke structure are given in Equation~\eqref{eq:ctlsem}, where $\sigma = (s_0, s_1, \dots)$ denotes a path.}
\begin{equation}
\label{eq:ctlsem}
\begin{aligned}
    & |[ p |]_M = \{s \in S : p \in L(s) \} \\
    & |[ \lnot \theta |]_M = S/|[\theta|]_M \\
    & |[ \theta_1 \lor \theta_2 |]_M = |[\theta_1|]_M \cup |[\theta_2|]_M
    \\
    & |[ \textbf{EX} \theta |]_M = \{s \in S : \exists t \in S \text{ such that } t \in |[\theta|]_M\ \text{ and } (s,t) \in R \}
    \\
    & |[ \textbf{EG} \theta |]_M = \{s \in S : \exists \sigma \text{ such that } s = s_0 \text{ and } \forall k \in \mathbb{N}, s_k \in |[\theta|]_M\}
    \\
    & |[ \textbf{EF} \theta |]_M = \{s \in S : \exists \sigma \text{ and }  \exists k \in \mathbb{N} \text{ such that } s = s_0 \text{ and } s_k \in |[\theta|]_M\}
    \\
    & |[\textbf{E} \theta_1 \textbf{U} \theta_2|]_M = \{s \in S: \exists \sigma \text{ and } \exists k \in \mathbb{N} \text{ such that } s = s_0 \text{ and }\\ && \mathllap{s_i \in |[\theta_1|]_M , s_j \in |[\theta_2|]_M \text{ for } i < k, j\geq k\}}
\end{aligned}
\end{equation}
\change{Note that we can get the semantic formulas for $\textbf{A}T$ for all the temporal operators, $T$, by replacing $\exists \sigma$ for $\forall \sigma$ (similarly replace $\exists t \in S$ for $\forall t \in S$) in the set definitions for the semantics.}

\change{Further, the operators \textbf{F} and \textbf{G} are obtained from \textbf{U} simply as $\textbf{F}\theta = \textit{true}\, \textbf{U}\, \theta$ and $\textbf{G}\theta = \neg \textbf{F} \neg\theta $, where $\textit{true}$ is the atomic proposition true in all states.}

It should be noted that CTL can be described with a subset of temporal and logical expressions as it is possible to create formulae from different terms. For example, the statements "there is no path such that eventually $\theta$ holds" and "for all paths $\lnot \theta$ always holds" are equivalent and specified in CTL by $\lnot \textbf{EF} \theta = \textbf{AG} (\lnot \theta)$.

\begin{figure}[ht]
\begin{minipage}{.4\textwidth}
    \centering

    \Description{The Kripke structure starts in the state s_0, which can only transition to state s_1. When in the s_1, the state can either loop and stay in s_1 or transition to the state s_2. Finally, when in s_2, the state can stay there by looping or transition to s_0.}
    \includegraphics[width=0.8\textwidth]{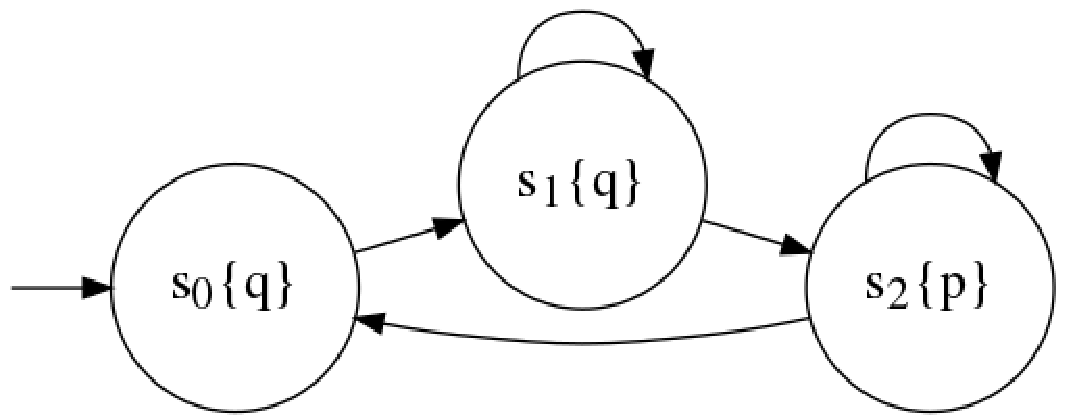}
    \caption{An example of a Kripke structure with $S = \{s_0, s_1, s_2\}$. 
    }
    \label{fig:model}
\end{minipage}%
\begin{minipage}{.6\textwidth}
    \centering
    \Description{A state transition tree of the previous figure. The root of the tree is s_0 and it transitions downwards to the valid states it can reach. The root and its child (s_1) are highlighted in yellow as the first proposition holds in those states. Then the right child (s_2) and its right child (s_2) are highlighted in blue as the second proposition holds in those states. All other nodes are not highlighted as these do not adhere to the CTL formula.}
    \includegraphics[width=0.6\textwidth]{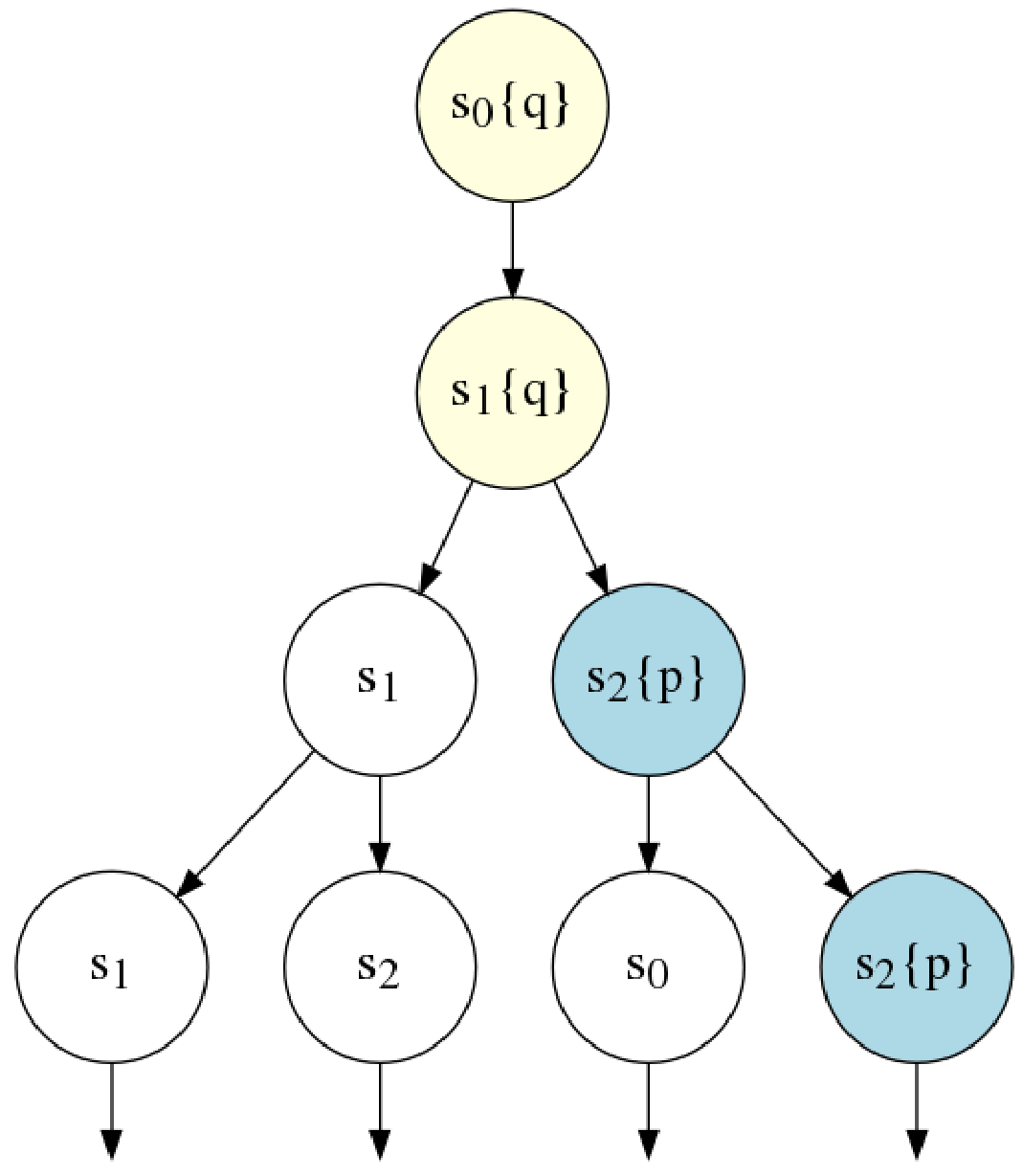}
    \caption{A computation tree for the automata given in Figure~\ref{fig:model} that shows the CTL operation $\textbf{E}(q \textbf{U}p)$. The proposition $q$ holds in states $s_0$ and $s_1$, but the proposition $p$ holds in state $s_2$.
    }
    \label{fig:ctltree}
\end{minipage}

\end{figure}

\begin{example}
Figure~\ref{fig:model} gives a model and the relative computation tree is given in Figure~\ref{fig:ctltree} with an example CTL operation.
\end{example}

The model checking problem is to find all valid states that satisfy a temporal logic formula. Alternatively, one could just ask whether the formula is true in the initial states. Given $M$ (a Kripke structure) and $\theta$ (a temporal logic formula), then find all $s \in S$ such that $M$ and $s$ models $\theta$, the semantics of which is denoted by \change{$M, s \models \theta \iff s \in |[ \theta |]_M$.}
\change{Model checking has been explored in usage for verifying quantum programs against extensions of temporal logics. This can be seen in Section~\ref{sec:qctl} and \ref{sec:lang:misc}.}

A reason for choosing model checking as a verification technique is that it efficiently searches over all possible states of a Kripke structure in a completely automated way. 
However, the main issue of model checking is the state explosion problem \cite{Clarke2012}.
Due to the design of the Kripke structure, an increase in the size of the system can increase the number of states in the structure massively. This can make it very difficult for the system to be verified quickly.
Over the last few decades a number of methods have been developed to address the state explosion problem. For example, bounded model checking (BMC) \cite{BMC} only considers finite computation trees, effectively meaning that the system is only checked up to a certain point in its temporal evolution.

Another technique is CounterExample-Guided Abstraction Refinement (CEGAR) \cite{CEGAR,MCabs}, which starts by creating an abstract model that {\em simulates} the original (concrete) model. The abstract system is then model checked against a given universally path-quantified temporal logic property. If the property is satisfied then we are done, since the abstract model encompasses all the possible behaviours of the concrete model. If instead a counterexample to the property is returned, then this is compared in the concrete model: if it is an actual counterexample, then the model checking fails since this is a ``real'' bug of the concrete model. Otherwise, the counterexample is spurious and the abstraction is refined so that the counterexample no longer fails inside the abstraction. The newly-obtained abstraction is then model checked again -- the process is repeated until the property is either verified or a concrete counterexample is found \cite{CEGAR}.

\subsection{Deductive Verification \label{sec:bg:deduc}}
For a full review of deductive verification the reader is referred to, \eg \cite{Huisman19}.
Unlike model checking, which exhaustively explores the possible states that software can be in, deductive verification formally verifies programs through logical inference. Further model checking can ensure certain properties about software automatically, whereas deductive verification can be used to verify complex properties about programs in a way that is understandable by humans. For example, a subroutine can be shown to always return a certain result no matter the input by using deductive verification.
The Floyd-Hoare logic \cite{Hoare69} is studied below as an example and will be useful later \change{(see Sections \ref{sec:qhl}, \ref{lang:QHLProver}).}

A program is given some preconditions, assumptions and rules that can be used for verification, and postconditions, goals or requirements to meet after the program has run with the given preconditions. A program is valid if the postconditions can be inferred from the given preconditions using inference rules.
This is often written in the form of a Hoare triple denoted by $\{P\}S\{Q\}$, where $P$ is a precondition, $Q$ is a postcondition and $S$ is a program statement. A Hoare triple is considered valid if a sequence of inference rules can be used to generate it. The basic inference rules are given in Equation~\eqref{eq:hoarerules}.
\begin{figure}[ht]
    \centering
    \begingroup
    \addtolength{\jot}{1em}
    \begin{align}
    \begin{aligned}
        &\frac{}
        {\assert{P} \textbf{ skip }\assert{P}}
        &\text{ (Skip)}\\
        &\frac{}
        {\assert{P[a/x]} x:=a \assert{P}} 
        &\text{ (Assign)}\\
        &\frac{\assert{P} S_1 \assert{Q}, \assert{Q} S_2 \assert{R}}
        {\assert{P} S_1; S_2 \assert{R}} 
        &\text{ (Composition)} \\
        &\frac{\hoare{b \land P}{S_1}{Q}, \hoare{\lnot b \land P}{S_2}{Q}}
        {\hoare{P}{\textbf{if }b\textbf{ then }S_1\textbf{ else }S_2}{Q}}
        &\text{ (Conditional)}\\
        &\frac{\hoare{P \land b}{S}{P}}
        {\hoare{P}{\textbf{while }b\textbf{ do } S}{\lnot b \land P}}
        &\text{ (While)}\\
        &\frac{P_2 \rightarrow P_1, \hoare{P_2}{S}{Q_2}, Q_2 \rightarrow Q_1}
        {\hoare{P_1}{S}{Q_1}}
        &\text{ (Consequence)}
    \end{aligned}
    \label{eq:hoarerules}
    \end{align}
    \endgroup
\end{figure}

The proof of a program or system can be created from the Hoare triple and the use of inference rules. These proofs are converted into proof obligations, which are mathematical formulae that are checked using one of a variety of software tools.
The most common tools for software verification are theorem provers and SMT (Satisfiability Modulo Theory) solvers. Theorem provers \cite{Wiedijk06} allow programmers to write the obligations that are to be met in a completely formal environment. Then lemmas and theorems about these obligations can be derived from definitions created within the tool. Normally, the process of proving an obligation is interactive and so the programmer will write the proof with assistance from the tool. \change{Theorem provers are used in a number of the tools discussed in Section~\ref{sec:vqpls}.}

In comparison, SMT solvers \cite{DeMoura09} convert obligations into logical formulae over a theory, such as, \eg the natural numbers, rationals and bit vectors. Alongside a statement the user wishes to assert, the solver can automatically check if the formulae are valid. If not, the solver can provide a counterexample. \change{In particular, one tool has made use of SMT solvers for verifying quantum programs; which we discuss in Section~\ref{lang:QBricks}.}

Deductive verification still suffers from scalability issues similar to model checking.
Unlike model checking, it requires programmers to have a deeper understanding of why the obligations are correct. This is both an advantage and a setback, since it can take a long time to prove complex obligations that could be solved automatically using model checking. However, it is possible to create human-readable proofs as to why a program is correct.

For examples of some of the most used theorem provers, the reader is referred to \cite{Wiedijk06}; an introduction to SMT solvers is given in \cite{DeMoura09} {and a deeper study is given in \mbox{\cite{SMTHandbook}}.}

\section{Formal Quantum Verification Methods \label{sec:qver}}
This section aims to introduce the theoretical ideas that have been used in pursuit of the verification of quantum programs. While this section covers some theories, it is not a complete list. Further theories include quantum \Markov chains \cite{Feng13} and quantum automata \cite{Kondacs97}, which are given a brief introduction in a previous survey \cite{Ying18} with further references therein.

\subsection{Quantum Weakest Precondition \label{sec:qwp}}
In deterministic programming, the weakest precondition \cite{Dijkstra} gives a method of transforming the problem of checking whether a program is valid in Hoare logic into a problem of determining whether a precondition implies the said weakest precondition.
More formally for deterministic programs, given a program $S$ and a predicate postcondition $Q$, then the weakest precondition $\text{wp}(S)(Q)$ is the precondition to $S$ such that for all preconditions $P$ with $\{P\}S\{Q\}$, then ${P\implies \text{wp}(S)(Q)}$.

Whilst a probabilistic version of the Hoare logic has been developed and can be used as a means to verify quantum programs \cite{Zuliani2000}, D'Hondt and Panangaden \cite{Panangaden06} demonstrated one can develop a quantum Hoare-style logic using density matrices. This then allows for the notion of a quantum weakest precondition.
The difference in definition is that now the program $S$ is a quantum program\change{, where $S(\rho)$ is the density matrix after applying program $S$ to density matrix $\rho$}; the \change{precondition $P$ and postcondition $Q$ are each a quantum predicate, which is a Hermitian operator with positive eigenvalues upper bounded by 1; and a valid precondition $P$ must satisfy $\tr(P\rho) \leq \tr(Q \; S(\rho))$ for all density matrices $\rho$.}
\change{We can write $\{P\}S\{Q\}$ if $P,Q$ and $S$ follow the final inequality.} Thus, the quantum weakest precondition $\text{wp}(S)(Q)$ is defined such that for all valid preconditions $P$, $\tr(P\rho) \leq \tr(\text{wp}(S)(Q)\rho))$ for all density matrices $\rho$.

With this notion, it is possible to change the verification problem of quantum programs to that of calculating quantum preconditions. {In a sense, the quantum weakest precondition gives the most ``general'' precondition for a postcondition, meaning that as long as we have a ``specific'' precondition we can always reach the same postcondition as the ``general'' precondition.
The quantum weakest precondition is a concept that can see usage in different verification systems depending on the language and design used. In Section~\mbox{\ref{sec:qhl}}, we will see an example of its usage.}

\subsection{Quantum Hoare Logic \label{sec:qhl}}
Ying \cite{Ying11} has been developing the Quantum Hoare Logic (QHL) over the last decade, as an extension of the standard Floyd-Hoare logic. Introduced in his original work, the classical while-language is extended to a quantum version and the Floyd-Hoare logic is amended to verify the extension. The quantum-while language is:
\begin{align*}
    S ::= & \textbf{ skip}
    \text{ | } q := 0
    \text{ | } \bar{q} := U \bar{q}
    \text{ | } S_1 ; S_2
    \text{ | } \textbf{measure } M[\bar{q}] : \bar{S}
    \text{ | } \textbf{while } M[\bar{q}] = 1 \textbf{ do } S.
\end{align*}

The commands do the following operations: \textbf{skip} does nothing, ${q := 0}$ initialises a qubit, $\bar{q} :=U \bar{q}$ performs a unitary operation on a number of qubits, ${S_1 ; S_2}$ is composition of statements, $\textbf{measure } M[\bar{q}] : \bar{S}$ measures some qubits and performs a program from $\bar{S}$ depending on the result and $\textbf{while } M[\bar{q}] = 1 \textbf{ do } S$ performs $S$ until a "false" measurement is read.


The quantum-while language stands out because it does not define programs in terms of describing quantum circuits. This allows for an imperative approach for writing quantum programs, rather than the very low-level idea of constructing a circuit. Another feature to highlight is the use of measurement within the language. The \textbf{measure} command replaces the classical \textbf{if} statement and the \textbf{while} statement requires measurement on a set of qubits during each iteration.

The Quantum Hoare Logic then extends the Hoare triple $\{P\} S\{Q\}$, \change{where $S$ is a program written in the quantum-while language and $P, Q$ are quantum predicates (as defined in Section~\ref{sec:qwp}). However, these predicates are additionally upper bounded by the identity operator, $I$, and lower bounded by the zero operator $0$. They are bounded in that for any predicate $P$ (used in QHL), then for all density matrices $\rho$ we have $\tr(0\rho) \leq \tr(P\rho) \leq \tr(I\rho)$, thus $0 \leq \tr(P\rho) \leq 1$.}

Inference rules can be used to create Hoare triples for quantum programs depending on the statement. Beyond that, the notion of weakest precondition can be used to generate valid Hoare triples. If a desired postcondition is wanted, then rules can be used to find the weakest precondition for a program.
The correctness of evaluating the Quantum Hoare Logic on a program and the condition that evaluation terminates (notions of partial and total correctness respectively) are given in \mbox{\cite{Ying11}}.

Work has continued on this logic over the last decade to further improve it and create an implementation, as will be seen in \change{Section~\ref{lang:QHLProver}}.
As an example of an improvement to be made on the grammar, it is clear to see that there is no classical functionality defined. This prevents some quantum algorithms, such as Shor's algorithm \cite{Shor97}, being fully implemented within the quantum-while language. A recent work \cite{Feng20} has extended the quantum-while language to include classical variables and the logic has been reworked to show the extension is verifiable.
Another version of Quantum Hoare logic has been proposed in \cite{Kakutani09}, which is not designed around the usage of weakest precondition. Further, this other version verifies Selinger's QPL \cite{selinger04}, a quantum programming language based on flowcharts and featuring classical bits as well as \textbf{if} statements.

\begin{example}
\label{ex:qhl}
Here we give an example of using Quantum Hoare logic and the quantum weakest precondition to verify Deutsch's algorithm. \change{Given $f:\{0,1\} \rightarrow \{0,1\}$}, recall that Deutsch's algorithm \change{determines} the value of $f(0) \oplus f(1)$ \change{with a single evaluation of $f$}. {It should be noted that this example does not make use of an ancillary qubit for the sake of simplicity.} We begin by writing Deutsch's algorithm in the quantum-while language:
\begin{equation*}
    \textit{Deutsch}=[q:=0; q:=Hq; q:=O_fq; q:=Hq; \textbf{measure } M[q] : \textbf{skip}, \textbf{skip}]
\end{equation*}

Here, $H$ is the single-qubit Hadamard gate and $O_f$ is the quantum oracle defined by the matrix
\begin{equation*}\begin{pmatrix} (-1)^{f(0)} & 0 \\ 0 & (-1)^{f(1)} \end{pmatrix}\end{equation*}
Further, the measurement operators $M$ consist of measurements on the computational basis:
\begin{equation*}
\Big\{
M_0 = \begin{pmatrix} 1 & 0 \\ 0 & 0 \end{pmatrix},
M_1 = \begin{pmatrix} 0 & 0 \\ 0 & 1 \end{pmatrix}
\Big\}
\end{equation*}

The result of receiving either measurement outcome in the \textbf{measurement} statement is to simply \textbf{skip}. Using the definition of the weakest precondition for Quantum Hoare Logic, we wish to find the weakest precondition of the postcondition $\textit{Post}=(1-f(0)\oplus f(1))\ket{0}\bra{0} + (f(0)\oplus f(1)) \ket{1}\bra{1}$. This postcondition states that $q$ should be in the appropriate state depending on the value of $f(0) \oplus f(1)$. Here, we will calculate the weakest precondition of the \textbf{measurement} statement and leave the rest of the calculation for the \textit{Deutsch} program in Appendix~\ref{app:hoare}.

From Proposition 7.1 in \cite{Ying11}, we have that
\begin{equation*}
    wp.(\textbf{measure } M[\bar{q}]:\bar{S}).P := \sum_m M^\dagger_m (wp.(S_m).P) M_m
\end{equation*}
and so for our case, we have
\begin{equation*}
    wp.(\textbf{measure } M[q]:\textbf{skip},\textbf{skip}).\textit{Post} := M^\dagger_0 (wp.(\textbf{skip}).\textit{Post}) M_0 + M^\dagger_1 (wp.(\textbf{skip}).\textit{Post}) M_1.
\end{equation*}
Further to this, the weakest precondition of \textbf{skip} is simply $wp.(\textbf{skip}).P = P$. Thus, we have that
\begin{align*}
    wp.(\textbf{measure } M[q]:\textbf{skip},\textbf{skip}).\textit{Post} & := M^\dagger_0 (\textit{Post}) M_0 + M^\dagger_1 (\textit{Post}) M_1 \\
    & = (1-f(0)\oplus f(1))\ket{0}\bra{0} + (f(0)\oplus f(1)) \ket{1}\bra{1}\\
    & = \textit{Post}
\end{align*}
giving the Hoare triple $\{\textit{Post}\} \textbf{measure } M[q]:\textbf{skip},\textbf{skip} \{\textit{Post}\}$.

Following the rules for quantum weakest precondition of the Quantum Hoare Logic,
the resulting Hoare triple for the entire program is $\{ I \} \textit{Deutsch} \{ \textit{Post} \}$, where $I$ is the single qubit identity operator.
This weakest precondition means that \textit{Deutsch} can have \change{any precondition and the program will always produce the correct result. This is because quantum predicates are upper bounded by $I$ and $I$ is the most general precondition we can have such that $Post$ is the postcondition.}
\end{example}

\subsection{Quantum Computation Tree Logic \label{sec:qctl}}
Various notions of extending Computation Tree Logic (CTL) to the quantum case have been studied, and implemented into model checking algorithms. For example, \cite{Feng13} investigated a quantum extension of probabilistic CTL (PCTL). Recently, in \cite{xu21} the authors created a different extension of CTL that uses the concept of fidelity (which measures how much a density matrix state is changed after being acted on by a super-operator).

Other temporal logics have been studied both as quantum extensions of the original logic \cite{Mateus09, Yu19} and how temporal logics can model behaviour in quantum systems\change{; examples of investigating linear temporal properties can be found in \cite{Bhatia20} and more general $\omega$-regular properties in \cite{Feng17}.}
In this section, the notion of quantum computation tree logic given in \cite{Baltazar08} is presented.

The Quantum Computation Tree Logic (QCTL) is a temporal logic used to reason about the behaviour of a quantum Kripke structure.
\change{Formally:
\begin{definition}
A (finite) quantum Kripke structure over a set of qubits $\text{qB}$ and variables $X$ is a tuple $(S, R)$ where:
\begin{itemize}
    \item $S \subset \mathcal{H}_{\text{qB}} \cross \mathbb{R}^X$ is a set of pairs $(\phi, \rho)$, where $\phi$ is a quantum state of the qubits $qB$ and $\rho$ is an assignment of variables to reals;
    \item $R \subseteq S \cross S$ is a relation such that for any $(\phi, \rho)$, there exists $(\phi', \rho')$ such that $((\phi, \rho), (\phi',\rho')) \in R$.
\end{itemize}
\end{definition}
Note that in comparison to the standard definition of Kripke structures, propositions are embedded as variables into the state rather than labels on specific states. In the literature for quantum model checking, names alternate between Kripke structures and Markov chains. These structures vary in definition, but will loosely follow the structure of a classical Kripke structure where each state can perform some transitions and has some labels (propositions) associated with it.
}

\change{By combining the decidable fragment of the exogenous quantum propositional logic (dEQPL) \cite{Chadha09} with the classical CTL used in model checking, we get QCTL.}
\change{The grammar for dEQPL used in \cite{Baltazar08} is:
\begin{align*}
    & \text{Classical Formulae} \\
    \alpha & ::= \bot_c
    \text{ | } \text{qb}
    \text{ | } \alpha \Rightarrow_c \alpha \\
    & \text{Terms} \\
    t & ::= x (\in Var)
    \text{ | } m (\in \mathbb{Z})
    \text{ | } (t + t)
    \text{ | } (tt)
    \text{ | } \text{Re}(\ket{\mathsf{T}}_A)
    \text{ | } \text{Im}(\ket{\mathsf{T}}_A)
    \text{ | }\textstyle{\int} \alpha \\
    & \text{Quantum Formulae} \\
    \gamma & ::= t \leq t
    \text{ | } \bot_q
    \text{ | } \gamma \Rightarrow_q \gamma.
\end{align*}
We distinguish between logic on classical formulae, $\alpha$, and quantum formulae, $\gamma$, by using subscript $c$ and $q$ respectively. Further, we can abbreviate other connectives ($\lnot, \land, \lor, \Leftrightarrow, \top$) using $\bot$ and $\Rightarrow$.
}

\change{Classical formulae describes the set of qubits we wish to measure from using classical logic statements and qubit symbols $\text{qb}$ from $\text{qB}$. Terms describe numerical expressions we can make with additional variables, constants, and functions for getting information about the quantum state.
The term $\text{Re}(\ket{\mathsf{T}}_A)$ denotes the real part of the amplitude of the quantum state from a subset of qubit symbols, $A \subseteq \text{qB}$ (similarly $\text{Im}(\ket{\mathsf{T}}_A)$ for the complex part).
The term $\textstyle{\int}\alpha$ denotes the probability that $\alpha$ holds when measuring all qubits.
Finally, quantum formulae allow us to reason about terms by using logical expressions and comparison formulae. This allows us to reason about the state of a quantum system at a specific time step.
}

In \cite{Baltazar08}, it is shown that QCTL is sound and (weakly) complete. Further, an algorithm was developed that checks if a QCTL formula is satisfiable by extending an algorithm used for model checking CTL. The grammar for QCTL is:
\begin{align*}
    \theta & ::= \gamma
    \text{ | } (\theta \Rightarrow_q \theta)
    \text{ | }  \textbf{EX}\theta
    \text{ | }  \textbf{AF}\theta
    \text{ | }  \textbf{E}[\theta\textbf{U}\theta]
    \\ & \text{where $\gamma$ is a dEQPL quantum formula.}
\end{align*}

\change{The semantics of the QCTL given above for the temporal operators are similar to CTL formula except they act over paths on a quantum Kripke structure.} Note that QCTL uses a subset of the temporal logic operations from CTL and the other temporal \change{operations} can be derived from this subset.


Unlike some of the other formal methods discussed in this section, this logic would be used in a similar way to the model checking described in Section~\ref{sec:bg:model}. This would involve converting properties of a quantum program or circuit into the QCTL language.
By providing a specification for the program, the QCTL formula can be checked for model satisfiability using the algorithm given in \cite{Baltazar08}. The other QCTLs introduced at the start of this section \cite{Feng13, xu21} found use in model checking quantum \Markov chains.

We now give two examples of QCTL \cite{Baltazar08} formulae.

\begin{example}
Denote $\Box \alpha $ as $(\textstyle{\int} \alpha) = 1$. This is a dEQPL formula that states a classical formula $\alpha$ holds with probability 1 after measuring all qubits. The following formula is used in \cite{Baltazar08} as the formula for verifying a single bit version of the BB84 protocol \cite{BB84}: 
\begin{equation*}
    B = (\Box (b_A \Leftrightarrow_c b_B)) \Rightarrow_q \textbf{A}[(\lnot_q (\Box e)) \textbf{U} ((\Box e) \land_q ((\Box k) \Leftrightarrow_q (\textstyle{\int} m = 1)))] .
\end{equation*}

The formula means "If Alice and Bob are in the same basis $(\Box (b_A \Leftrightarrow_c b_B))$, then down all (quantum) paths the protocol has not ended ($\lnot_q (\Box e))$) until it has ended $(\Box e)$ and the generated key is equal to the value of the qubit used ($(\Box k) \Leftrightarrow_q (\textstyle{\int} m = 1)$)". 
\end{example}

\begin{example}
As a second example, consider Deutsch's algorithm. Denote $b = f(0) \oplus f(1)$ as a classical bit. Further, use $m$ as a classical bit to say when the algorithm has performed the measurement operation and let $q$ denote the qubit used for the algorithm. \change{Usage of $m$ is required to specify temporal behaviour for Deutsch's algorithm.}
We then have the following formula:
\begin{equation*}
    D = \textbf{A}[(\lnot_q (\Box m)) \textbf{U} ((\Box m) \land_q ((\Box b) \Leftrightarrow_q (\textstyle{\int} q = 1)))] .
\end{equation*}

The formula reads "for all (quantum) paths ($\textbf{A}$), the qubit is not measured ($\lnot_q (\Box m)$) until it is measured ($\Box m$) and the measured qubit gives us the correct result with certainty (${(\Box b) \Leftrightarrow_q (\textstyle{\int} q = 1)}$)".
\change{For the $Deutsch$ program given in Example~\ref{ex:qhl} to satisfy this specification, it would need to initialise new qubits $m$ and $b$ at the start of the program and only modify $m$ after the rest of the program has been completed. Then the program could be modelled by a quantum Kripke structure and shown to satisfy $D$.}
\end{example}

\subsection{Path Sums \label{sec:pathsum}}
Path sums are a representation of unitary operators in terms of a summation of exponential polynomials with different quantum states. This representation was used in \cite{Amy19} to show the equivalence of quantum programs.
Equation~\eqref{eq:psums} shows a path sum represented as a unitary operator. Note that $\mathbf{x} = (x_1, \dots , x_n)$ is a collection of Boolean variables or constants (the input signature), $P$ is a (phase) polynomial with inputs $\mathbf{x}$ and $\mathbf{y}$, and $f : \mathbb{Z}_2^n \times \mathbb{Z}_2^m \rightarrow \mathbb{Z}_2^n$ is a multi-variable Boolean function (the output signature):
\begin{equation} \label{eq:psums}
        U : \ket{\mathbf{x}} \rightarrow \frac{1}{\sqrt{2^m}} \sum_{\mathbf{y} \in \mathbb{Z}_2^m} e^{2 \pi i P(\mathbf{x},\mathbf{y})} \ket{f(\mathbf{x},\mathbf{y})}.
\end{equation}

Path sums are used to represent the semantics of Clifford$+R_k$ circuits\footnote{See Appendix~\ref{app:cliff} for information on Clifford circuits.} for a fixed $k$.
Reduction rules can be applied to decrease the size of the circuit and so it is easy to check the equivalence of quantum circuits.
Experiments in \cite{Amy19} have shown that Clifford$+T$ circuits with a large number of qubits and gates can be efficiently and automatically verified.
This gives a useful representation to efficiently verify quantum circuits. However, it will require support through a translation tool to be used to verify circuits written in a high-level programming language, which could include classical components. \change{An extension of path sums has been used for verifying a quantum programming language. This usage is discussed in Section \ref{lang:QBricks}.}

\begin{example}
The path sum representation of the Pauli gates are listed below:
\begin{align*}
X & : \ket{x} \rightarrow \ket{1-x} \\
Y & : \ket{x} \rightarrow e^{2\pi i \frac{(2x + 1)}{4}}\ket{1-x} \\
Z & : \ket{x} \rightarrow e^{2\pi i \frac{x}{2}}\ket{x}
\end{align*}
\end{example}


\subsection{The ZX-Calculus}

The ZX-calculus \cite{Coecke11} \change{can be utilised as} an alternate form of verification. The ZX-calculus is a graphical tool designed to convert quantum circuits into a graphical model and back again (but not all graphs are quantum circuits). These graphical models or networks model wires in the form of edges and some operations in the form of vertices.

\begin{figure}
    \centering
    \hfill
    \begin{subfigure}[b]{0.3\textwidth}
    \centering
    \includegraphics[width=0.1\textwidth]{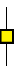}
	\begin{equation*}
	    \frac{1}{\sqrt{2}} \begin{pmatrix} 1 & 1 \\ 1 & -1 \end{pmatrix}
    \end{equation*}
    \end{subfigure}
    \hfill
    \begin{subfigure}[b]{0.3\textwidth}
    \centering
    \includegraphics[width=.15\textwidth]{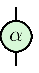}
    \begin{equation*}
	    \begin{pmatrix} 1 & 0 \\ 0 & e^{i\alpha} \end{pmatrix}
    \end{equation*}
    \end{subfigure}
    \hfill
    \begin{subfigure}[b]{0.3\textwidth}
    \centering
    \includegraphics[width=.15\textwidth]{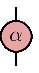}
    \begin{equation*}
	    e^{\frac{i\alpha}{2}}\begin{pmatrix} \cos{\frac{\alpha}{2}} & -i\sin{\frac{\alpha}{2}} \\
	    -i\sin{\frac{\alpha}{2}} & \cos{\frac{\alpha}{2}} \end{pmatrix}
    \end{equation*}
    \end{subfigure}
    \caption{The Hadamard node (left), green spider (middle) and red spider (right) with their matrix representation. The vertices can have more than one edge enter and exit them, but for simplicity we consider the case when only one edge enters a vertex. The spiders represent rotations on an axis, with the green spider rotating around the $Z$-axis and the red spider rotating around the $X$-axis. Note that when $\alpha = 0$, the spiders are simply identity operators and when $\alpha = \pi$ the spiders represent the pauli-$Z$ and pauli-$X$ gate respectively.}
    \Description{The Hadamard node is a yellow square. The green spider node is a green circle containing an alpha symbol inside it. Similarly, the red spider node is a red circle with the same alpha symbol. All the nodes have two wires (lines) entering from the northern and southern points of the nodes.}
    \label{fig:zx}
\end{figure}

The ZX-calculus consists of three main vertices: the Hadamard node, green spiders and red spiders (shown in Figure~\mbox{\ref{fig:zx}}). Using a set of rewrite rules, it is possible to add or remove various vertices.
This allows the creation of optimised circuits and comparison between circuits.

While the ZX-calculus is a useful tool for \change{low-level verification}, it is an example in highlighting the difference in what the other tools described are trying to achieve. The ZX-calculus is useful for optimising circuits and showing the equivalence of circuits. It can be used to verify properties of simple circuits, such as the teleportation protocol \cite{Coecke11}.

\change{However, recently, the ZX-calculus has also been used to verify some properties of oracle-based algorithms \cite{Carette2021}.
As research continues into the ZX-calculus, the calculus may reach a point where it can be used to represent programs. The Scalable ZX-calculus \cite{Carette2019} can represent circuits with a parameterised number of qubits and is used in \cite{Carette2021} to represent quantum algorithms.
The ZX-calculus has several other applications\footnote{Such as usage in circuit optimisation, surface codes and measurement-based quantum computation.} and a much more in-depth introduction can be found in \cite{deWetering2020}.}

\begin{example}
\change{Here we first give a few rewrite rules from \cite{Backens17} in Figure~\ref{fig:rewrite}. Whilst many different rewrite rules can be used, these are a few key ones for this example.

\begin{figure}
    \centering
    \includegraphics[width=.5\textwidth]{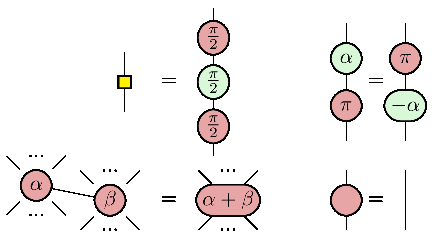}
    \caption{A number of different rewrite rules \cite{Backens17}. It is important to note that the colours of spiders can be swapped (so green spiders replace red spiders and vice versa). The summation of scalars is modulo $2\pi$.}
    \label{fig:rewrite}
    \Description{
    Rule 1: Hadamard nodes are the same as a concatenation of a red, green then red spiders (each with angle pi/2).
    
    Rule 2: A green spider (angle a) followed by a red spider (angle pi) is the same as a red spider (pi) followed by a green spider (-a).
    
    Rule 3: If two spiders, with angles a and b, are joined by a wire, then they can be merged into a single spider with angle a + b (mod 2*pi)
    
    Rule 4: A spider with no angle (angle 0) is the same as just having a wire.
    }
\end{figure}

Using these rules, it is now possible to show optimisations of different circuits. For example, we can use the rewrite rules to show that applying two Hadamard gates to a circuit is the same as doing nothing (performing the identity operation). This derivation is given in Figure~\ref{fig:derive}.

\begin{figure}
    \centering
    \includegraphics[width=.8\textwidth]{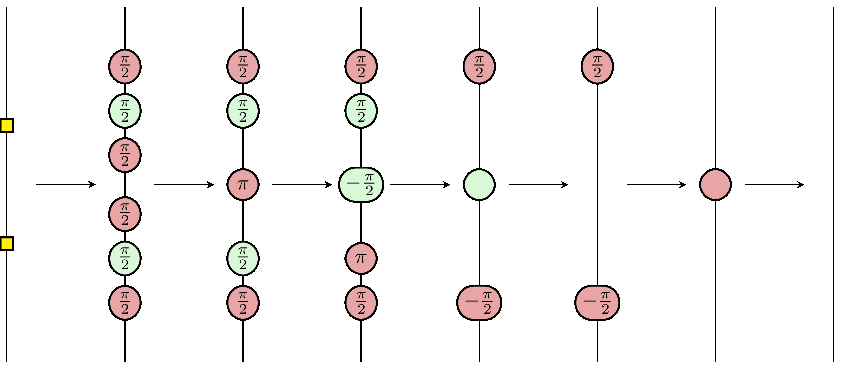}
    \caption{Derivation of two Hadamard gates being equivalent to the identity operation using the rules of the ZX-calculus. The derivation uses instances of the rewrite rules given in Figure~\ref{fig:rewrite}. Firstly, the Hadamard gates are decomposed and then the red spiders are merged. Two of the spiders are swapped and then the green spiders are merged. This combined green spider can be removed by the identity rule and the final red spiders can be similarly merged and removed. \change{An alternative way of proving the Hadamard gate is self-inverse can be found in \cite{Backens17}.}}
    \label{fig:derive}
    \Description{
    Let H denote the Hadamard node, R(a) red spiders with angle a and similarly G(a) for green spiders. Further "-" denotes wires outside of spider arguments (where they represent minus). We have that:
    
    (-H-H-) = (-R(pi/2)-G(pi/2)-R(pi/2)-R(pi/2)-G(pi/2)-R(pi/2)-) = (-R(pi/2)-G(pi/2)-R(pi)-G(pi/2)-R(pi/2)-) = (-R(pi/2)-G(pi/2)-G(-pi/2)-R(pi)-R(pi/2)-) = (-R(pi/2)-G(0)-R(-pi/2)-) = (-R(pi/2)---R(-pi/2)-) = (--R(0)--) = (-----)
    }
\end{figure}}
\end{example}

\section{Design of Verification Frameworks and Quantum Programming Languages \label{sec:criteria}}
In this section, several trade-offs and properties that are desired from a programming language for quantum verification are given. Whilst the trade-offs are similar to what classical theorem provers need to consider, different requirements need to be met for the language due to the nature of quantum computation. These criteria will later be used to highlight the differences of the available verification frameworks for quantum programming.

\subsection{Trade-offs}

\paragraph{Environment}
This concerns the environment in which the programmer creates their programs. There are a few options available when considering this. Firstly, the language could be embedded within an available theorem prover. This gives the benefits of the host language, as well as access to libraries and community support. Further, someone familiar with the environment would be able to pick up the language fairly easily. However, embedding within a theorem prover does mean that the quantum verifier also suffers from the limitations of the said theorem prover.

An alternative is to create a new environment dedicated to the verification of quantum programs from scratch. This gives more freedom in being able to meet the specification the designer creates. On the downside, it may take longer to develop than other methods. {It will take longer to fully build a dedicated tool for the verification of programs versus building off or extending a well-known tool}.

Another approach is to extend a quantum programming language with a verification framework. This would allow both programs to be verified and executed on a simulator or quantum hardware. However, the choice of language needs to be considered and the framework would need to be updated whenever the underlying language gets updated.

\paragraph{Interactive vs Automated}
Here, one needs to consider whether the user should interact with the proofs they are constructing and how much should a tool automate the process of generating the proof. One can design a fully interactive prover, a fully automated prover or a mix between the two.

If the framework is designed to be interactive, then it may take more man-hours to construct proofs. However, this brings the benefit of the user obtaining a better understanding as to how a program is (or is not) valid.
On the other hand, an automated-focused framework could only require the push of a button to prove programs, possibly leaving the user in the dark as to why the program is correct.

{The size of systems requiring verification are a factor in which type of tool to use. A large system might not be suitable for an interactive approach, requiring the use of an automated approach. In \mbox{\cite{Miller10}}, the authors highlight the use of model checkers and SMT solvers for the verification of avionics systems. The case studies provided contain systems with reachable state spaces of up to $1.5 \times 10^{37}$ states.}

\paragraph{Executability and Separability}
Another property to consider is what should be executable within the framework. This will depend on the environment chosen for the framework and how mixed the specification is with the programs created.
If a program is to be run on quantum hardware, the specification and verification of a program need to be separated from its definition. This is because the verification of quantum programs occurs classically (at least for now) whereas programs would be executed on quantum hardware. This puts the separation between the definition and execution of programs and proofs as a high priority.


Running programs on a simulator is quite different. The reason for running a quantum program on a simulator is to check the outcomes of the program. With a verification framework, this becomes unnecessary. Thus, the separation between program and proof could be avoided if this approach is taken.


 
\subsection{Limitations of Quantum Programs}
Due to the difference in nature between quantum and classical computing, quantum programming languages have different requirements to classical ones. In \cite{Huang19} a discussion on a number of bugs to avoid when implementing a quantum programming language is given. Here a few of the ideas presented in \cite{Huang19} are discussed as are some other thoughts. The paper goes into more details on common bugs that programmers may create due to human error, but this is omitted.

\paragraph{No-cloning}
One of the key properties to meet is that the no-cloning theorem \cite{Wootters82} is adhered to. The rules that quantum mechanics follows do not allow for the copying of arbitrary quantum data. Therefore, any quantum programming language should have inbuilt functionality to forbid (perfect) cloning. For example, if programs and proofs are mixed within the language, then the no-cloning theorem can be \change{proved} within the language.
\change{Alternatively, no-cloning can be achieved by making it built into the language through the use of types or linear logic.}

\paragraph{Limited Classical Functionality}
A useful approach of designing a quantum computer is by having a classical computer being able to access a small quantum processor, known as quantum random access memory (QRAM), in a similar way to a graphics processing unit (GPU) \cite{Knill96}.
The quantum part of the computer should perform purely quantum operations, having very little classical functionality within it. Because of this restriction, quantum programming languages need to be designed so that there is very limited classical functionality within the language as to avoid affecting the quantum system.

It should be noted though that there are some nice features that can come with having classical functionality. For instance, oracles implement classical functions into a quantum circuit and having classical functionality would allow for easy implementation of oracles. Silq \cite{Bichsel20} implements this in an easy-to-do way. Striking the right balance of how much classical functionality to have is therefore very important to consider.

Depending on the amount of classical functionality, there are two ways to verify a program with quantum and classical components. One method is to combine a quantum verifier with a classical verifier, each handling the verification of separate parts of the program. The alternative would be for the verifier to have suitable logic to handle classical and quantum functionality together.
\change{\paragraph{Program Parameters}
Moving away from low-level quantum circuit description languages and into the realm of quantum programs requires the use of parameters. Using parameters allows quantum programmers to easily describe their program in a very general manner and then provide specific values during runtime. For example, Grover's algorithm can use parameters to describe the size of the circuit and a general oracle function that it can take as input. This provides complexity from a verification perspective as the more parameterised a program is, the harder it is to verify for correctness or other properties.
}

\change{
For some quantum programs, it can be useful to use the output of measurement results to influence the control of a program. This concept is a form of \textit{dynamical lifting}, which allows classical data to be used to affect the control of a quantum program. Classical data can be used as a parameter before or during runtime. Types of control flow include determining the number of wires to use, running different circuits after a qubit is measured (\eg the teleportation protocol) or a notification to redo the computation (\eg repeat until success loops). Recent extensions of Quipper have shown how dynamic lifting can be implemented in practice \cite{Fu2022, Colledan2022}. Techniques such as dynamic lifting will require specific methods to be handled by verifiers.
}

\paragraph{Ancillary Cleaning}
Ancillaries are qubits that are introduced into the system in an initial state, used for a computation temporarily and then returned to their initial state. This process of returning the ancillary to its initial state is known as cleaning and is important for programmers or language designers to take into account. In classical computation, ancillary bits can be easily removed automatically by the processor {through garbage collection, since classical bits can be discarded without affecting the state of the program.}
However, if ancillary qubits are not cleaned, there is the potential for measurement outcomes to be affected {due to entanglement between the main qubits and the ancillary ones}. It should be decided by the designers whether programs should automatically uncompute ancillaries or if the programmer should perform this task. 

A key feature in the programming language Silq \cite{Bichsel20} is that ancillaries are automatically uncomputed. Whether this feature will be seen in future verifiable programming languages is uncertain. From a verification perspective it would be advantageous to verify properties of ancillary qubits. For example, one such property is verifying that an ancillary qubit returns to its initial state so that it can be removed from the computation.
\change{\paragraph{Types for Quantum Variables}
Another issue that is faced by designers of quantum programming languages is the design of types within their language. This includes problems such as whether measuring a quantum variable causes it to remain the same type, and what properties of a quantum variable need to be embedded into a type.
}

\change{
Another factor to consider when handling types is how to detect entanglement between different variables. Twist \cite{Yuan2022} features a unique typing system that allows the programmer to change the type of a variable depending on whether it is entangled with other variables or not. This is known as purity checking and it is handled through operations that are run before and during the running of a program. The techniques developed in Twist can be of use for future languages, but advancements need to be made to perform full purity checking before the program is run. Purity checking can be considered a form of verification and is a problem that needs to be explored with different tools.
}

\paragraph{Algorithm Milestones}
Designers of quantum languages often show the capabilities of their language by implementing a quantum algorithm. One should think of each algorithm as a milestone that should be reached. The lowest milestone to reach is being able to implement the Deutsch-Jozsa algorithm \cite{Deutsch1992} or the {quantum teleportation protocol \mbox{ \cite{Teleport}}} as these are fairly well-known algorithms. The next milestone would be to write either Grover's \cite{Grover96} or the Quantum Phase Estimation algorithm (many of these algorithms mentioned can be found in \cite{Nielsen11}).
This is because both involve some form of iteration and are slightly more complex than the Deutsch-Jozsa  and teleportation algorithms. The hardest algorithms to implement would be some of the 7 algorithms implemented in Quipper \cite{Green13}, which includes quantum walks on Binary Welded Trees \cite{WeldedTree} and the Ground State Estimation algorithm \cite{GroundStateEst}.

For verifiable quantum programming languages, it is also necessary to be able to prove these programs run correctly. Many quantum programming languages are already able to implement several of the common ``textbook'' algorithms and a variety of algorithms in other fields, such as quantum chemistry \cite{QChem}. So far, verifiable languages have only been able to prove about "textbook" algorithms. Whether they can prove facts about more advanced algorithms is yet to be seen. In Section~\ref{sec:alg} we discuss two non-textbook algorithms and their challenges to verification.

\section{Verifiable Quantum Programming Languages
\label{sec:vqpls}}

Here various quantum programming languages that can verify programs are discussed, highlighting their trade-offs, where they excel and their limitations. At the end of this section, other quantum verification tools, whose focus is not on the formal verification of programs, are briefly discussed.

\subsection{SQIR (and QWire) \label{lang:SQIR}}
The languages QWire~\cite{Paykin17}
and SQIR (Small Quantum Intermediate Representation)~\cite{Hietala20}
are domain specific languages built in the Coq interactive theorem prover~\cite{Coq20}.
QWire was one of the first quantum programming languages to be released with verifiable programs, while SQIR is a more recent language that has various improvements over QWire. These improvements include shorter code, better handling of ill-typed programs and the separation of semantics for unitary and non-unitary (\eg measurement) operations.

SQIR uses the various functionalities of Coq to act as a proof assistant for writing proofs about quantum programs. In order to verify a program, firstly the program is defined using a dedicated type, with qubits being referred to by a numerical value. Programs can be defined by one of two types: 
\textbf{base\_ucom}, 
which only contains unitary operations; or 
\textbf{com}, 
which allows for measurement.
Classical subroutines cannot be performed within a SQIR program, but it is possible to generate circuits using classical parameters.

Theorems can then be conjectured about the program. Unitary SQIR uses state vectors as part of the semantics, whereas full SQIR extends this to density matrices. The user proves the theorem with the assistance of the Coq framework.
This format separates programs from how they are specified/proved.
No-cloning is satisfied through the use of unitary gates and measurement being the only operations allowed in the language.

An example program and proof can be seen in Figure~\ref{fig:sqir}. Whilst the example given is quite trivial, it is possible to make generalisations about what can be proven. The creators of SQIR have already proven some properties about Grover's algorithm, notably the probability of measuring a marked element after $T$ steps being $\sin^2((2T+1) \arcsin(\sqrt{k/2^n}))$ (where $n$ is the number of qubits and $k$ is the number of marked elements).

\begin{figure}[ht]
    \centering
    \usemintedstyle{emacs}
    \begin{minted}[]{coq}
    Definition Program : base_ucom 2 := H 0; X 0; CNOT 0 1; X 0.
    
    Local Open Scope R_scope.
    
    Definition xor_state : Matrix (2^2) (1^2) : = 
    1/\sqrt{2} .* (|0,1> .+ |1, 0>).
    
    Theorem Prog_Correct: uc_eval Program x (|0,0>) = xor_state.
    Proof.
      intros
      unfold Program; simpl.
      unfold xor_state; simpl.
      autorewrite with eval_db; simpl; try lia.
      solve_matrix.
    Qed.
    \end{minted}
    \Description{The SQIR program has two definitions and a theorem. The definitions are Program, which describes the two qubit quantum circuit that is being used, and xor-state, which defines the expected output state. The theorem, Prog-Correct, says that if the Program evaluates two qubits in the zero state, then the resulting state is the xor-state. The theorem also has a proof that this result actually hold.}
    \caption{A simple program and proof written in SQIR, which transforms the state $\ket{00}$ into $\frac{1}{\sqrt{2}}(\ket{01} + \ket{10})$.
    }
    \label{fig:sqir}
\end{figure}

Coq requires interaction from the user for theorem proving, but there is some automation when reducing matrices {as seen in the ${solve\_matrix}$ function given in Figure~\mbox{\ref{fig:sqir}} (${solve\_matrix}$ is a tactic that attempts to simplify a matrix equality using a variety of sub-tactics).}
Further, SQIR benefits from a number of gates already implemented and verified within the language.
With its capabilities, SQIR has been able to prove properties about most textbook algorithms, including Grover's, the Quantum Phase Estimation and Shor's algorithms.

The key issue that may hinder SQIR is that circuits require a predefined number of qubits. Should ancillary qubits be required in the circuit, the user must define these at the start, rather than introducing them when required.

\subsection{QHLProver \label{lang:QHLProver}}


The Quantum Hoare Logic, as described previously in Section~\ref{sec:qhl}, has been implemented into the Isabelle/HOL proofing tool~\cite{Liu19}. This implementation uses a slightly simpler version of the quantum-while language, omitting the initialisation term $q:=0$.

\begin{figure}[ht]
    \centering
    \includegraphics[width=0.35\textwidth]{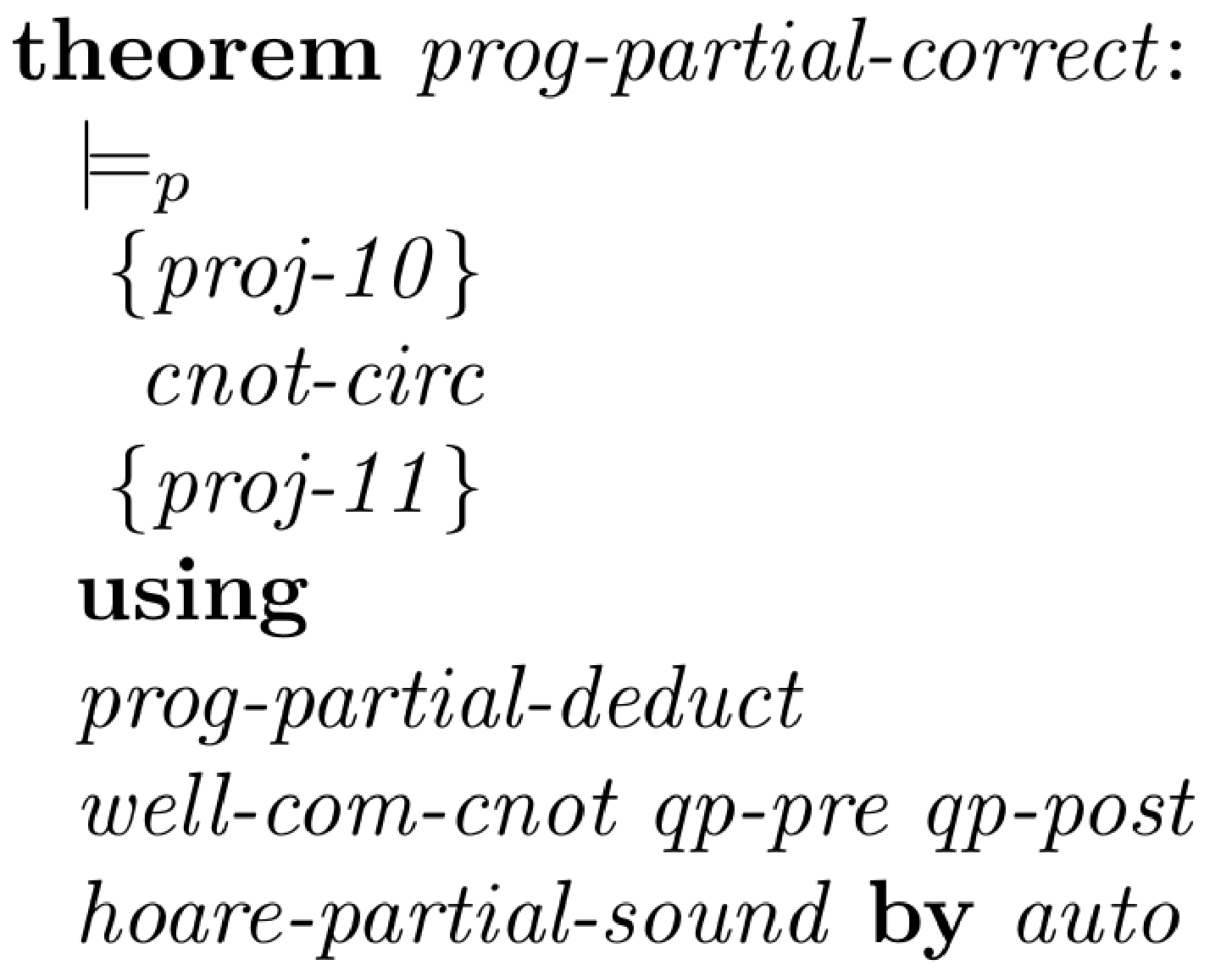}
    \Description{An Isabelle theorem titled prog-partial-correct. This theorem states the partial correctness of a CNOT operation with initial quantum state |10> evaluating to the qubit state |11>. This is proven using the lemmas described in the caption and the "by auto" command.}
    \caption{Documentation of a simple program that implements a single CNOT gate and checks that the state $\ket{10}$ is transformed to $\ket{11}$.
    \change{The state $\ket{10}$ is represented by the predicate \textit{proj-10} $= \ketbra{10}$ and a similar predicate represents $\ket{11}$.}
    The lemmas \textit{qp-pre} and \textit{qp-post} prove that the predicates are valid, \textit{well-com-cnot} confirms that the program is a valid and \textit{prog-partial-deduct} proves the deduction.
    {It would be possible to prove more general properties about CNOT, such as its effect on a general quantum state, but this requires experience with Isabelle and the QHLProver library. The full Isabelle theory file for this example is available in this repository: \mbox{\url{https://github.com/marco-lewis/QHLProverCNOT}}}}
    \label{fig:qhl:cnot}
\end{figure}

A full documentation of the implementation is available in the Isabelle Archive of Formal Proofs (AFP)~\cite{QHLProver-AFP}. This implementation is referred to as QHLProver. By using Isabelle, verification is not fully automated, but some automation is used to make manual proving easier when handling complex matrices, which are used to define gates and oracles. 
An example can be seen in Figure~\ref{fig:qhl:cnot}, showing how a program is validated within the Isabelle framework.

QHLProver is similar to SQIR in a few ways. The way programs are proved is akin to that of Coq, where programs, states 
and density matrices 
are defined and then theorems can be conjectured about them. It also suffers the flaw of SQIR where programs require a predefined number of qubits and so ancillary qubits need to be defined at the start of computation.

Programs are written using a specific type (com), which encodes the terms of quantum-while, and the predicates used within the pre- and post-conditions are density matrices. These all need to be verified for a specific triple to be correct as seen in Figure~\ref{fig:qhl:cnot}.

Unlike SQIR, which defines programs as quantum circuits, QHLProver uses an extension of the quantum-while language to define programs. Whilst there is no theory written for no-cloning within the documentation, no-cloning is adhered to due to the simple grammar that only allows for unitary operations and measurement.

As mentioned previously, there is no classical functionality in the quantum-while language and this is reflected within the Isabelle implementation. Classical parameters are used to extend gates to operate on a subset of qubits. Constructing oracles for different algorithms is done by writing functions defined in terms of natural numbers to booleans, which are then used to create a complex matrix from the matrix indexes.

QHLProver suffers from a lack of implemented and verified logic gates. Currently, the language implements the Pauli gates, Hadamard gates and some gates used within Grover's algorithm. Out of the common quantum gates, only the Hadamard gate is verified. This makes it difficult to create new programs as the user must verify simple gates (such as CNOT).

\paragraph{CoqQ}
\change{A recent tool called CoqQ~\cite{CoqQ} builds on the work from QHLProver. It shares a number of similarities, such as using QHL and the quantum-while language. As in the name, CoqQ uses the Coq theorem prover instead of Isabelle/HOL.}

\change{CoqQ enhances the semantics of QHL by using improved inference rules and allowing dynamic initialisation of qubits. This allows the framework to verify non-textbook algorithms such as algorithms for solving the hidden subgroup problem~\cite{Nielsen11} and the hidden linear function problem~\cite{BGK}; and the HHL algorithm (see Section~\ref{sec:alg:hhl}).}






\subsection{Isabelle Marries Dirac \label{lang:IMD}}

\begin{figure}[ht]
    \centering
    \includegraphics[width=0.6\textwidth]{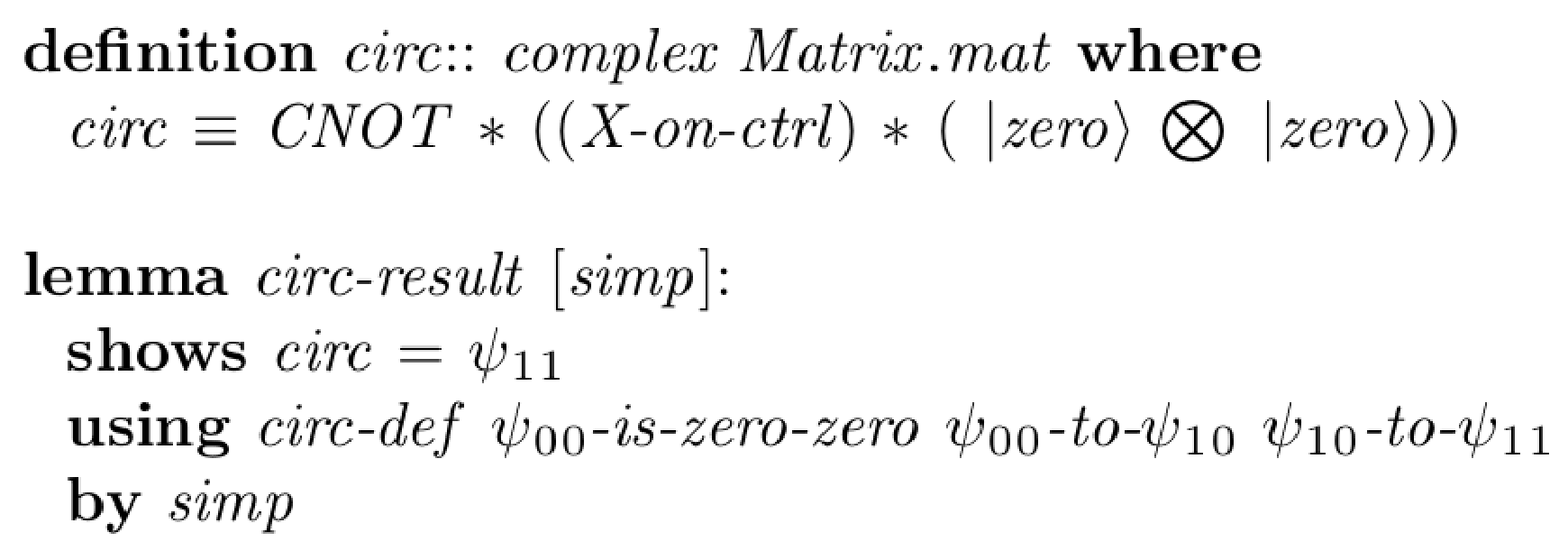}
    \Description{The circuit circ is defined as a complex matrix. It is defined as a NOT gate on the first qubit followed by a CNOT gate. This is then used in the lemma circ-result, which says that circ = |11>. This is proven using the definition of circ, an assertion that |0>x|0> = |00> and the evolution of the quantum state (from |00> to |10> to |11>).}
    \caption{Documentation of a simple CNOT program similar to that of QHLProver. The circuit performs a NOT-gate on the first qubit and then the CNOT gate with the first qubit as control. The lemma \textit{circ-result} uses the circuit definition and how the state evolves through each transition to show the resulting state.
    The full Isabelle theory file for this example is available in this repository: \url{https://github.com/marco-lewis/IMD_CNOT}.}
    \label{fig:imd}
\end{figure}

A recent work~\cite{Bordg2020} provides another instance of verified quantum computing using the Isabelle theorem prover. Unlike QHLProver's use of the quantum-while language and Hoare logic, Isabelle Marries Dirac (IMD) uses the standard matrix formalisation approach of quantum computing to prove properties about algorithms and protocols. Because of this, IMD is closer to a verifiable mathematical library, rather than a verifiable programming language. {An example can be seen in Figure~\mbox{\ref{fig:imd}}.}
%

Firstly, matrices are defined to have a fixed size, whether through a variable $n$ or a value. Ancillaries are needed to be defined by the user and taken into account when defining matrices.
Fortunately, programs in IMD are defined using the dot and Kronecker product of matrices. This makes it easier to add in ancillary definitions.

Since IMD is a mathematical library, classical functionality is possible so long as proofs are developed in the Isabelle theorem prover. Oracles are constructed in a similar way to QHLProver, where the matrix value at an index is determined by a function with indexes as input.
Various properties about measurement are implemented within the library such as the probability of a given outcome.
It is possible to prove properties about not just standard algorithms (\eg the Deutsch-Jozsa algorithm) but also different quantum information theoretic results such as quantum teleportation. \change{One result to mention in particular is that IMD explicitly proves the no-cloning theorem within Isabelle. The other tools within this section do not provide such a proof in the framework they use, but the definitions of states and operations that the tools use allow for no-cloning to be followed.}
More results are discussed in~\cite{Bordg2020}.


Further, Echenim and Mhalla~\cite{Echenim2021} have used IMD and density matrices from QHLProver to prove properties about projective measurements (measuring qubits in a basis different from the computational basis $\{\ket{0}, \ket{1}\}$). They continued to develop more information theory results by proving the CHSH inequality~\cite{CHSH}. 
As can be seen, the featured libraries have more proofs relating to mathematical concepts of quantum computing that cannot be implemented in a programming language.
The documentation for IMD and the extension~\cite{Echenim2021} are available in the Archive of Formal Proofs~\cite{Isabelle_Marries_Dirac-AFP, Projective_Measurements-AFP}.

\subsection{QBricks \label{lang:QBricks}}
QBricks~\cite{Chareton20}
is a circuit-based verifiable quantum programming language built in the Why3 framework.
The language is purposefully built so that the writing of programs is separated from the specification to be proved about the program.
Programs are written using QBricks-DSL, which is a domain specific language, and the specifications are written in QBricks-SPEC. {Figure~\mbox{\ref{fig:qbrcks}} gives the specification and definition for the oracle used in the Deutsch-Jozsa algorithm.\footnote{At the time of writing, a tutorial article is being written by the team.}}

\begin{figure}[ht]
    \centering
    \usemintedstyle{emacs}
    \begin{minted}[]{coq}
    val function deutsch_oracle (f: bitvec -> int)(n:int): circuit
      requires{1<=n}
      requires{(not (constant_bin f n)) -> balanced_bin f n}
      ensures{width result = n+1}
      ensures{forall x: bitvec. forall  y: matrix complex.  is_a_ket_l y 1 ->
      path_sem result (kronecker (bv_to_ket x) y) = 
        kronecker (bv_to_ket x)  (xor_qbits (ket 1 (f x)) y)}
    \end{minted}
    \Description{See https://github.com/Qbricks/qbricks.github.io/blob/main/Case_studies/deutsch-jozsa.mlw for the implementation of deutsch_oracle.}
    \caption{The definition of a Deutsch-Jozsa oracle within the QBricks language~\cite{Chareton20}. The \textit{require} statements note the preconditions of the oracle (there is at least 1 qubit and if $f$ is not constant then it is balanced). The first \textit{ensures} statement maintains the width of the oracle from input to output. The second \textit{ensures} statement gives the usual definition of a quantum oracle ($O\ket{x}\ket{y} = \ket{x}\ket{f(x) \oplus y}$) from necessary preconditions. This function can be used later to prove properties about the Deutsch-Jozsa algorithm. Full documentation is available at \url{https://github.com/Qbricks/qbricks.github.io}.}
    \label{fig:qbrcks}
\end{figure}

The verification process of programs uses the ideas of the weakest precondition, path sums and quantum Hoare logic to generate proof obligations, building on previous works to suit the requirements of the language. These obligations are then proved using automatic SMT solvers such as Alt-Ergo~\cite{AltErgo} and Z3~\cite{DeMoura08}.
QBricks still requires some interactivity from the user when writing the specification, but as mentioned these are mostly proved automatically.

The language features a vast array of functionalities, including the capability of introducing ancillary qubits in the middle of code, unlike SQIR and QHLProver. The framework adheres to the no-cloning theorem as it only allows certain unitary operations within its DSL.

QBricks still has some limitations though. Currently, there are no built-in capabilities to measure qubits within QBricks-DSL.
Further, QBricks is not designed to interact with classical data, but classical parameters can be used in the generation of circuits.

Despite these limitations, so far QBricks has been able to verify properties for the Phase Estimation and Shor's algorithms, which are the most complex algorithms verified by languages so far.


\subsection{Related Verification Tools \label{sec:lang:misc}}
\paragraph{Quantum Model Checking}
\change{The Quantum Program/Protocol Model Checker (QPMC)~\cite{QPMC} is a tool developed to check a quantum extension of probabilistic CTL (qCTL; different from the QCTL discussed in Section~\ref{sec:qctl}) against programs that are modelled by quantum \Markov chains. This allows programmers to use model checking techniques against quantum programs. However, QPMC is limited in that one needs to write their program as a \Markov chain for it to be checked. This can be circumvented by a tool known as Entang$\lambda$e~\cite{Anticoli16, Anticoli18}, which allows quantum programmers to convert Quipper programs into \Markov chains that can then be checked against qCTL properties.
}

\paragraph{Feynmann - Path Sums}
\change{Using the path sum technique from Section~\ref{sec:pathsum}, a Haskell library known as Feynman was produced to perform simulation, verification and equivalence checking of quantum circuits \cite{Amy19, AmyThesis}.}

\paragraph{Equivalence Checking using Binary Decision Diagrams (BDDs)}
\change{Another form of equivalence checking has been developed through the use of BDDs and extensions. There are two notable tools that take this direction. The first approach~\cite{Burgholzer2021, Wille2022} makes use of an extension of binary decision diagrams called quantum multiple-valued decision diagrams (QMDDs)~\cite{Niemann2017}. The second approach SliQEC~\cite{Chen2022, Wei2022} uses a bit-slicing technique to represent complex numbers efficiently in standard BDDs~\cite{Tsai2021}.
}

\paragraph{PyZX}
This tool is a module in the Python language that implements the ZX-calculus~\cite{Kissinger20}.
Still currently under development, PyZX is able to convert back and forth between circuits and ZX graphs. This allows circuits to be simplified and optimised using the rules of the ZX-calculus.
PyZX is not designed for reasoning about programs, but is useful for checking the equivalence of circuits.

\paragraph{CertiQ}
This is a verification framework developed for Qiskit~\cite{Qiskit} that verifies if a compiled circuit is the same as the circuit that is programmed by the user~\cite{Yunong19}.
Interestingly, it makes use of Z3 and other SMT solvers to perform this verification automatically, requiring the user to only input a few lines of specification. While this uses automatic verification, again it should be highlighted that this is used for verifying circuit equivalence, similarly to PyZX, and not for reasoning about programs.

\paragraph{QSharpCheck}
This tool extends Q\# with a means of testing programs~\cite{Honarvar20}.
Users can initialise qubits, notably their phase, and a number of different postconditions to be met by the resulting qubits. With this, the user can then define parameters that are used to run the tests. When run, test cases are randomly generated, executed on the program and checked they meet the postconditions given. This makes it very easy to quickly test a few properties of a program. However, this tool is designed for testing purposes and not for formal verification of programs. Despite this, it may find use for simpler programs that do not need to be verified extensively\footnote{In early 2021, Microsoft added some testing and debugging functionality to Q\#: \url{https://docs.microsoft.com/en-us/azure/quantum/user-guide/testing-debugging?tabs=tabid-vs2019}}.

\section{Verifying Complex Quantum Algorithms
\label{sec:alg}}
Whilst many standard quantum algorithms have been verified in a number of the tools described above, specific techniques will be needed to verify complex algorithms. Here we study the problems that arise for verifying the Harrow-Hassidim-Lloyd algorithm \cite{HHL} and the Binary Welded Tree quantum walk algorithm \cite{WeldedTree}.

\subsection{The Harrow-Hassidim-Lloyd Algorithm \label{sec:alg:hhl}}

\begin{figure}
    \centering
    \includegraphics[width=0.8\textwidth]{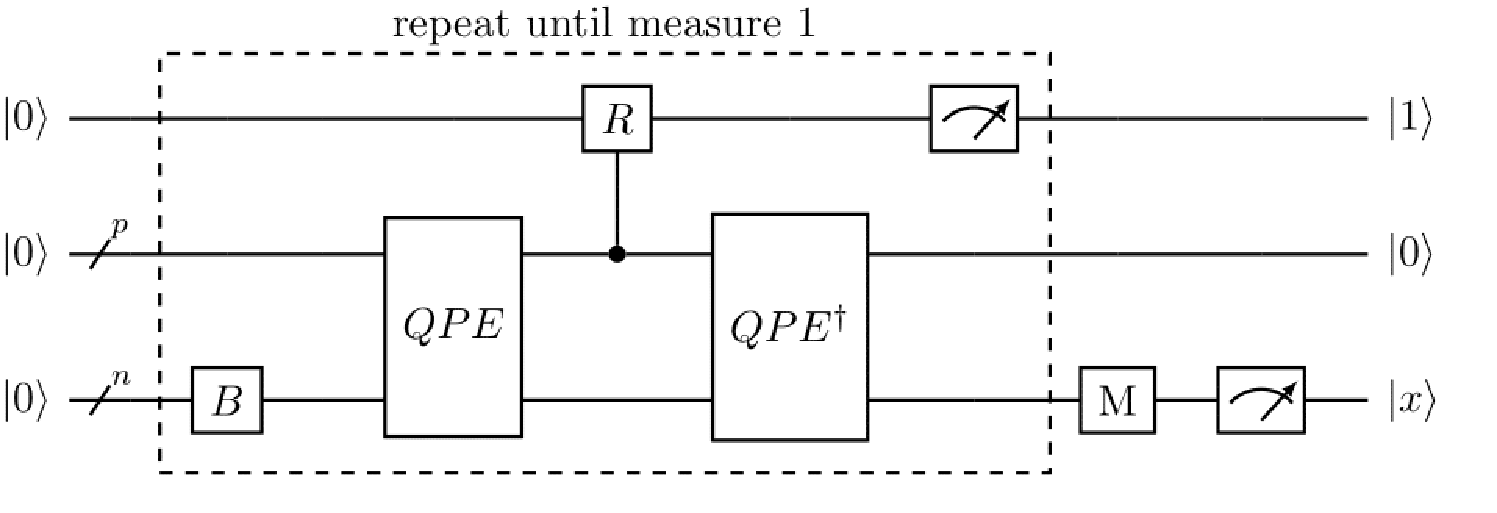}
    \Description{The circuit begins with 3 quantum registers: the ancillary of 1 qubit, the estimation register of p qubits, and the vector register of size n. Each register is initialised to 0. The B gate is applied to the vector register, then a QPE gate is applied to the vector and estimation register. Using the estimation register as a control, an rotation gate R is applied to the ancillary qubit. After that, the QPE routine is reversed and the ancillary qubit is then measured. These steps are all contained in a box that repeats until a 1 is measured, which then applies an M gate and measurement on the vector register. This results in the final state 10x (where x is the final vector).}
    \caption{The quantum circuit for the HHL algorithm.}
    \label{fig:hhl}
\end{figure}

The Harrow-Hassidim-Lloyd (HHL) \cite{HHL}  quantum algorithm solves linear systems of equations. Specifically, given a sparse, Hermitian matrix $\mathbf{A}$ and unit vector $\mathbf{b}$, find $\mathbf{x}$ such that $\mathbf{Ax} = \mathbf{b}$.\footnote{Although it is possible to change a linear system problem that uses a non-Hermitian matrix to one that uses a Hermitian matrix (details are within \cite{HHL}).} 
{Using a matrix $\mathbf{A}$ as described above, the classical Conjugate-Gradient method can be used to find $\mathbf{x}$ in time $O(N\kappa s \log{(1/\epsilon)})$, where $N$ is the size of $\mathbf{A}$, $\kappa$ is the condition number of $\mathbf{A}$, $s$ denotes how sparse $\mathbf{A}$ is and $\epsilon$ is the error \mbox{\cite{ConjGrad}}. In the case of the HHL algorithm, the actual output is an approximation of $\mathbf{x}$ using a measurement matrix $\mathbf{M}$. This approximation can be found classically in $O(N \kappa \text{poly}(1/\epsilon))$. The run time of HHL is $O(\log{(N)} \kappa^2 s^2 / \epsilon^3)$ and returns the answer with high probability, providing an exponential speedup with respect to $N$.}
The circuit for the HHL algorithm is given in Figure~\ref{fig:hhl} and some Silq code is presented in Figure~\ref{alg:hhl}.

\begin{figure}
    \centering
    \usemintedstyle{vs}
    \begin{minted}[tabsize=2,breaklines]{d}
        // Init b
        b_q := orB(b_q);

        // QPE
        eigen_qs := H_n[prec](eigen_qs);
        b_q := qpe_ham_loop[n, prec](eigen_qs, t, hamU, b_q);
        eigen_qs := reverse(qft[prec])(eigen_qs);
        
        // Controlled rotation
        anc := control_rot[n, prec](eigen_qs, approx_lambda, t/(2*pi), anc);

        // Inverse QPE
        eigen_qs := qft[prec](eigen_qs);
        b_q := qpe_ham_loop[n, prec](eigen_qs, t, revhamU, b_q);
        eigen_qs := H_n[prec](eigen_qs);
        
        // Measure ancillary
        repeat_loop = not(measure(anc));
    \end{minted}
    \caption{Main loop body for the HHL algorithm written in Silq. Full code available at \url{https://github.com/marco-lewis/silq-hhl}.}
    \Description{A sample of source code from silq-hhl/hhl.slq.}
    \label{alg:hhl}
\end{figure}

Here we give a brief explanation of how the HHL algorithm works. Let $N = 2^n$ and the $N$-length unit vector $\mathbf{b}$ be represented by the quantum state $\ket{b} = \sum_{i= 0}^{N-1} \mathbf{b}_i \ket{i}$. This is loaded into a $n$-qubit register using an operator $B$. \Hamiltonian simulation is used to represent $\mathbf{A}$ in a quantum phase estimation call. {The unitary matrix $U$ in quantum phase estimation is $e^{-i\mathbf{A}t}$ for HHL and rotates $\ket{b} = \sum_{j} \beta_j \ket{\mu_j}$ around the eigenvectors of $\mathbf{A}$, which are $\mathbf{\mu_j}$ and $\ket{\mu_j}$ being their quantum state representation (similarly to $\ket{b}$).}

{Performing the quantum phase estimation ($QPE$) call entangles the eigenvalue representations of $\mathbf{A}$ in a new quantum register with their associated eigenvectors. This gives a quantum state of the form $\sum_j \ket{\lambda_j}\ket{\mu_j}$.} The eigenvalues are then embedded into the phase of the quantum state using controlled rotations on an ancillary qubit (using an operator $R$), leaving the quantum state as $\sum_j \ket{\lambda_j}\ket{\mu_j}(C\ket{0} + \frac{1}{\lambda_j}\ket{1}$).

The quantum phase estimation routine is undone and the ancillary qubit is then measured. If it returns $\ket{0}$, then the entire quantum state is dumped and the algorithm is run again. If it returns $\ket{1}$, then the register that previously contained $\ket{b}$ now contains {$\ket{x} = \sum_j \frac{\beta_j}{\lambda_j} \ket{\mu_j}$}, a representation of $\mathbf{x}$. This can then be measured after applying $\mathbf{M}$, which changes the basis that $\ket{x}$ is measured in. 

The HHL algorithm has a number of features that will make formal verification of an implementation challenging. Some of these features are not common in the standard algorithms and are discussed.

\paragraph{Repeat until Success}
One of the key aspects involved in this algorithm is the Repeat until Success loop that is dependent on the measurement of an ancillary qubit.
This feature is also an aspect of Shor's algorithm \cite{Shor97}, since there are possibilities for a measured result to be invalidated through classical checks. 
{QHLProver is the only suitable verification framework that can handle \mbox{Repeat-until-Success} loops dependent on measurement outcomes. This is achieved through the verification of the \textbf{while} statement in the quantum-while language.}

\paragraph{Subroutines}
As mentioned the HHL algorithm features the quantum phase estimation algorithm. This would need to be verifiable first before much progress could be made on the HHL algorithm. However, it is also important to consider what properties of quantum phase estimation need to be proved for different algorithms.
{For HHL, it will be important to show that $\mathbf{b}$ has a representation under the eigenvectors of $\mathbf{A}$. 
It will also be important to consider how the eigenvalue representations affects the phase of the quantum state. Verification will be needed here to prevent side effects within a program.
}

\paragraph{\Hamiltonian Simulation Approximation}
\Hamiltonian simulation is used to give a unitary approximation of the evolution of a \Hamiltonian system. This technique has been used with matrices of different types for different purposes (as is the case in HHL). A verification tool will need to ensure that an implementation of an approximation is actually a good approximation of the behaviour that is expected.
{Techniques for verifying that implementations of a \Hamiltonian are correct have not yet been studied (to the best of our knowledge). None of the tools given have verified an example of \Hamiltonian simulation. Tools embedded within theorem provers might be able to perform this verification but this will require separate study.}


\subsection{Walking on Binary Welded Trees}

The problem statement of the Binary Welded Tree (BWT) algorithm \cite{WeldedTree} is that there are two trees, of the same depth, that are joined together by a weld at their leaves. The task is to walk from one root node (the entrance node) to the root node on the other tree (the exit node). When at a node, you only have local knowledge, where you cannot ask where you are on the tree but you can identify edges you have been down. {For trees of depth $n$, one can walk to the weld and then walk down edges randomly until the exit node is reached. This is referred to as a random walk and has a worst case run time of $O(2^n)$, as one may visit almost every node in the tree.\footnote{One can achieve a polynomial classical algorithm, but this uses global knowledge of the walkers position on the graph.} The random quantum approach allows the walker to traverse down all edges from a node by entering a superposition of staying at the current node and traversing to a new node. This quantum approach achieves an exponential speed up in comparison to the classical approach, running in $O(\text{poly}(n))$ and the quantum system will collapse to the root of the other tree with high probability.}

\begin{figure}
    \centering
    \includegraphics[width=0.7\textwidth]{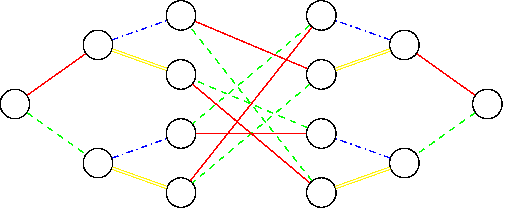}
    \caption{Examples of a binary welded tree with 4 colours used to colour the edges \change{(solid is red, dashed is greed, dot-dashed is blue and double-line is yellow)}.}
    \Description{Two binary trees of depth 3 are joined through a weld. The edges from the root to its children are coloured red and green. Then the edges to the leaves are coloured blue and yellow. The weld between the leaves of the tree are coloured red and green. The colouring given ensures that there are no two edges with the same colour entering a node.}
    \label{fig:weld}
\end{figure}

Each leaf node in the weld is connected to two leaves on the other tree. Further, the weld is designed such that it is cyclic and every leaf is in the cycle. An example of a binary welded tree is given in Figure~\ref{fig:weld}. To identify edges at a node, they are coloured. Edges only need to have a unique colour to the other edges on attached nodes. There is no constraint on the number of colours used, but it is possible to colour a tree using only 4 colours (this can be seen in Figure~\ref{fig:weld}).

Again, the reader should refer to \cite{WeldedTree} for full details on how the algorithm works. To explain briefly, a quantum register is used to represent the nodes and the register is initialised to the start node. The algorithm features a number of colour oracles that modify some ancillary register. A colour oracle returns a connected node if a node has an edge of that colour or the error state if not. Further, an additional qubit is used to flag the error state.



    

\begin{figure}
    \centering
    \usemintedstyle{vs}
    \begin{minted}[tabsize=2,breaklines]{d}
    def Ham[n:!N](const G_or: !N x uint[n]->lifted uint[n], t:!R, node_reg:uint[n], neighbour_anc:uint[n], err_anc:B){
      
      for col in [0..numOfCols){
        // Perform first V_c oracle
        [neighbour_anc, err_anc] := V(G_or, col, node_reg, neighbour_anc, err_anc);
    
        // Simulation of T operation
        [node_reg, neighbour_anc, err_anc] := sim_T(t, node_reg, neighbour_anc, err_anc);
    
        // Perform second V_c oracle
        [neighbour_anc, err_anc] := V(G_or, col, node_reg, neighbour_anc, err_anc);
      }
    
      return (node_reg, neighbour_anc, err_anc);
    }
    \end{minted}
    \caption{A section of Silq code \cite{Bichsel20} for the BWT Algorithm. The colours are looped over, $V$ represents the colour oracle function and $sim\_T$ represents the \Hamiltonian simulation that takes places. The colour oracle is used twice to uncompute the ancillaries (full code available at \url{https://github.com/marco-lewis/silq-binary-welded-trees}).}
    \Description{A sample of source code from silq-binary-welded-trees/BWTAlgorithm.slq.}
    \label{alg:bwt}
\end{figure}

\begin{figure}
     \centering
     \hfill
     \begin{subfigure}[b]{0.3\textwidth}
         \centering
         \includegraphics[width=0.3\textwidth]{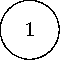}
         \caption{Initial state starting at the root node of the left tree}
         \label{fig:evo1}
         \Description{A root node with a 1 in.}
     \end{subfigure}
     \hfill
     \begin{subfigure}[b]{0.3\textwidth}
         \centering
         \includegraphics[width=0.3\textwidth]{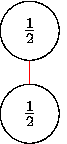}
         \caption{Evolution with red oracle}
         \label{fig:evo2}
         \Description{A root node and a child node with a red edge, each with 1/2 written in.}
     \end{subfigure}
     \hfill
     \begin{subfigure}[b]{0.3\textwidth}
         \centering
         \includegraphics[width=.9\textwidth]{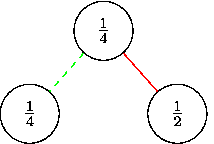}
         \caption{Evolution with green oracle}
         \label{fig:evo3}
         \Description{A new child node is added to the root node with a green edge. The root node now has 1/4 written in as does this new node.}
     \end{subfigure} \\
     \hfill
     \begin{subfigure}[b]{0.45\textwidth}
         \centering
         \includegraphics[width=0.7\textwidth]{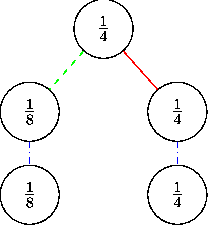}
         \caption{Evolution with blue oracle}
         \label{fig:evo4}
         \Description{New child nodes are formed from the children of the root node, each joined by a blue edge. These take half of the values from their parent nodes.}
     \end{subfigure}
     \hfill
     \begin{subfigure}[b]{0.45\textwidth}
         \centering
         \includegraphics[width=0.9\textwidth]{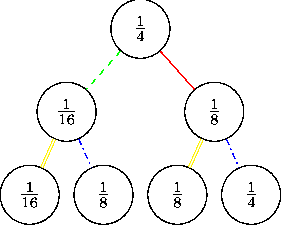}
         \caption{Evolution with yellow oracle}
         \label{fig:evo5}
         \Description{New child nodes are formed from the children of the root node, each joined by a yellow edge. Again these take half of the value from their parent nodes. The final tree is a root node (1/4), two children of the root (1/8 and 1/16) and their children (1/4, 1/8, 1/8 and 1/16 respectively).}
     \end{subfigure}
     \caption{Example of the first iteration performing the main procedure through each colour oracle. Inside each node is the probability of that node being measured. A variable can be set to change the proportion of probability moved to a connected node. In this example, half of the probability from a node is shared with its neighbour. Changing the order the oracles are called in would change the evolution of the system.}
     \label{fig:bwtevo}
\end{figure}

For a specific colour, the quantum register is put through the colour oracle. Then a rotation occurs through a \Hamiltonian simulation. The entangled node register and ancillary register are rotated so that the connected node is slightly rotated into the quantum state. Then the colour oracle is reversed.
The algorithm loops through the colour oracles and repeats this loop a specified number of times. By selecting an appropriate time, the algorithm will stop and return the label of the root of the right tree with high probability. Figure~\ref{fig:bwtevo} shows a diagrammatic evolution of the quantum state for the first iteration through the colour oracles.


Through the description of the algorithm, one can see that there are various features that will be difficult to formally verify. Some of these features are discussed.

\paragraph{Loop Invariants}
Unlike the measurement based loops in the HHL algorithm, the BWT algorithm features two classical loops: one for the colour oracles and one for repeating the process. {This leads to very deep circuits being created and so making the verification process slow from simply unwinding the entire loop.}
To verify these loops, either loop invariants need to be made or Bounded Model Checking could be used to unwind the procedure for a number of iterations.
{QBricks already features an $iter$ statement for repeating a quantum circuit a finite number of items. Such a feature could be used to verify aspects of BWT, but this is expected to be a challenging task.}



\paragraph{Quantum Objects - Nodes vs Qubits}
Whereas most of the standard algorithms use qubits to represent values, the BWT algorithm works on a ``quantum'' graph. While physically this graph would be represented by  qubits, it would be helpful for verification to verify about the nodes and graph structure easily. QuantumOne \cite{QuantumOne} is a recent endeavour into verifying about quantum objects rather than qubits.

\paragraph{Oracle Implementation}
In standard algorithms (\eg Deutsch-Jozsa), oracles only refer to a single function with defined constraints. However the oracles for the BWT algorithm are more complex, depending on a classical variable from a loop and featuring an error flag. This provides difficulty in terms of defining the expected behaviour for the oracle function.

\section{Conclusion \label{sec:con}}

This paper provides the reader with some insights into the current state of research for verification within quantum programming.
It should also demonstrate the importance of verification techniques keeping pace with the development of quantum hardware and software.


{Given the various topics discussed, we list a few open problems. Whilst this is not an exhaustive list, it should give room for researchers to investigate problems within formal verification of quantum programs:}
\begin{itemize}
    \item Many of the tools listed can only perform verification up to a certain extent. For example, QBricks is unable to handle measurement of qubits and QHLProver is unable to initialise qubits in the middle of a program. Further, most tools do not have the means to verify classical data and operations. So, can a verifier, such as QBricks or a new tool, be used to verify programs with measurement and classical bits? Can properties of various verifiers be merged together or are there practical limits in terms of complexity?
    \item Languages in the NISQ-era have seen development and it would be worth investigating how theoretical frameworks can be applied to them. Some tools, such as CertiQ, already provide some verification for programming languages. Can a tool be made for the formal verification of programs written in a widely used language (or a subset of it), such as Q\# \mbox{\cite{Svore18}}?
    \item We have seen examples of the complex quantum algorithms that are currently available. The question remains as to whether simple examples of these algorithms can be verified using tools currently available. \change{CoqQ~\cite{CoqQ}, mentioned at the end of Section~\ref{lang:QHLProver}, has been able to verify a few complex algorithms using QHL, but other techniques still need to tackle this hurdle.}
    \item Further, whilst general quantum programs have been discussed, a question remains over the verification of specific quantum techniques. For example, how to verify simulations of \Hamiltonians or quantum loops with classical control?
\end{itemize}

A final point to highlight is that verifiable programming languages need to be developed to be easy to learn or use. 
Non-verifiable languages have the benefit that users can quickly learn the syntax of the language and have access to dedicated documentation.
Those that require verification also require the user to learn the underlying tool that is used to write them if they have not used theorem provers before, e.g. Coq for SQIR and Isabelle for QHL. Making a tool that is easy to use and quick to learn would encourage more programmers to use them.

\begin{acks}
This work was supported by the Engineering and Physical Sciences Research Council (EPSRC project reference EP/T517914/1).
\end{acks}

\bibliographystyle{ACM-Reference-Format}
\bibliography{ref}

\appendix
\section{Clifford Gates \label{app:cliff}}

The Clifford gates \cite{Gottesman98} are the following gate operations:
\begin{equation}
\begin{matrix}
    H = \frac{1}{\sqrt{2}}\begin{pmatrix}
    1 & 1 \\
    1 & -1
    \end{pmatrix} &
    S = \begin{pmatrix}
    1 & 0 \\
    0 & i
    \end{pmatrix} &
    CX = \begin{pmatrix}
    1 & 0 & 0 & 0 \\
    0 & 1 & 0 & 0 \\
    0 & 0 & 0 & 1 \\
    0 & 0 & 1 & 0 
    \end{pmatrix}
\end{matrix}
\end{equation}
Circuits that can be generated from Clifford gates are referred to as being in the Clifford group and can be simulated efficiently on a classical computer, as stated by the Gottesman-Knill theorem \cite{Gottesman98}. It is unknown if all Non-Clifford circuits (i.e. circuits containing operations that cannot be broken down to Clifford operations) can be simulated efficiently.
If this was the case, then quantum computers could be efficiently simulated on classical computers.

The operation $S$ can be generalised to $R_k$, where
\begin{equation}
    R_k = \begin{pmatrix}
    1 & 0 \\
    0 & e^{\frac{2\pi i}{2^k}}
    \end{pmatrix}
\end{equation}
and it can easily be seen that if $k\geq 2$, then $(R_k)^{2^{k-2}} = S$ (with $R_2 = S$). For $k \geq 3$, $R_k$ is a non-Clifford operation. The circuits generated by using Clifford gates and the $R_k$ gate (for a fixed $k$) are referred to as the Clifford$+R_k$ group.
In particular, denote $T = R_3 = 
\begin{pmatrix}
1 & 0 \\
0 & e^{\pi i/4}
\end{pmatrix}$ and then $\text{Clifford}+R_3 = \text{Clifford}+T$. Algorithms for simulating Clifford$+T$ gates are discussed in \cite{Bravyi16}. The algorithms' runtimes are exponential based on the number of $T$ gates.

\section{Quantum Hoare Logic Example \label{app:hoare}}
As seen earlier, we have that $\{\textit{Post}\} \textbf{measure } M[q]:\textbf{skip},\textbf{skip} \{\textit{Post}\}$. We now calculate $wp.(q:=Hq).\textit{Post}$. For ease of notation, denote $c = 1-f(0)\oplus f(1)$ and $b = f(0)\oplus f(1)$ (so $\textit{Post} = (c\ket{0}\bra{0} + b \ket{1}\bra{1}$). Using the weakest precondition rules from Proposition 7.1 in \cite{Ying11}, we have
\begin{align*}
    wp.(q:=Hq).\textit{Post} & = H^\dagger \textit{Post } H \\
    & = H (c\ket{0}\bra{0} + b \ket{1}\bra{1}) H \\
    & = c\ket{+}\bra{+} + b \ket{-}\bra{-}
\end{align*}

and so we have the Hoare triple $\{c\ket{+}\bra{+} + b \ket{-}\bra{-}\} q:=Hq \{\textit{Post}\}$. Next we find the weakest precondition of $q := O_f q$ for $\{c\ket{+}\bra{+} + b \ket{-}\bra{-}\}$. Using the same rule we have
\begin{align*}
    wp.(q:=O_f q). (c\ket{+}\bra{+} + b \ket{-}\bra{-}) &= O_f^\dagger (c\ket{+}\bra{+} + b \ket{-}\bra{-}) O_f \\
    &= O_f (c\ket{+}\bra{+} + b \ket{-}\bra{-}) O_f \\
    &= ((-1)^{f(0)} \ket{0}\bra{0} + (-1)^{f(1)} \ket{1}\bra{1}) \\
    & (c\ket{+}\bra{+} + b \ket{-}\bra{-}) \\
    & ((-1)^{f(0)} \ket{0}\bra{0} + (-1)^{f(1)} \ket{1}\bra{1}) \\
    &= c ((-1)^{f(0) + f(0)} \ket{0}\bra{0} + (-1)^{f(0) + f(1)} \ket{0}\bra{1} \\
    & + (-1)^{f(1) + f(0)} \ket{1}\bra{0} + (-1)^{f(1) + f(1)} \ket{1}\bra{1}) \\
    & + b ((-1)^{f(0) + f(0)} \ket{0}\bra{0} - (-1)^{f(0) + f(1)} \ket{0}\bra{1} \\
    & - (-1)^{f(1) + f(0)} \ket{1}\bra{0} + (-1)^{f(1) + f(1)} \ket{1}\bra{1}) \\
    & \change{ = (c+b)(-1)^{2f(0)} \ketbra{0} } \\
    & \change{ + (c - b)(-1)^{f(0)+f(1)} \ketbra{0}{1} } \\
    & \change{ + (c - b)(-1)^{f(1)+f(0)} \ketbra{1}{0} } \\
    & \change{ + (c+b)(-1)^{2f(1)}\ketbra{1} } \\
    & = \ket{0}\bra{0} + \ket{0}\bra{1} + \ket{1}\bra{0} + \ket{1} \bra{1} = \ket{+}\bra{+}.
\end{align*}

The penultimate equality holds \change{because ${c - b = 1-2(f(x) \oplus f(y)) = (-1)^{f(x)+f(y)}}$.}
Thus, we have the Hoare triple $\{\ket{+}\bra{+}\}q:=O_f q\{c\ket{+}\bra{+} + b \ket{-}\bra{-}\}$.

It is not hard to see that $\{\ket{0}\bra{0}\}q:=Hq\{\ket{+}\bra{+}\}$ with $\ket{0}\bra{0}$ being the weakest precondition of the given statement and postcondition.

Finally, we can use the weakest precondition rule for initialisation:
\begin{equation*}
    wp.(q:=0).P=\ket{0}_q\bra{0}P\ket{0}_q\bra{0}+\ket{1}_q\bra{0}P\ket{0}_q\bra{1}
\end{equation*}

For simplicity, we can drop the $q$ notation from the equation, since we only have a 1-qubit program. Thus, it is easy to see that
\begin{align*}
        wp.(q:=0).\ket{0}\bra{0} &= \ket{0}\bra{0}\ket{0}\bra{0}\ket{0}\bra{0}+\ket{1}\bra{0}\ket{0}\bra{0}\ket{0}\bra{1} \\
        &= \ket{0} \bra{0} + \ket{1} \bra{1} = I
\end{align*}
and therefore the Hoare triple $\{I\}q:=0\{\ket{0}\bra{0}\}$ holds.
By following the sequential rule, it is easy to see that $wp.(\textit{Deutsch}).\textit{Post} = I$ and we get the desired Hoare triple of $\{I\}\textit{Deutsch}\{\textit{Post}\}$.


\end{document}